\documentclass[%
 aip,
 cha,%
 amsmath,amssymb,
 reprint,%
]{revtex4-1}

\usepackage[english]{babel}
\usepackage[utf8]{inputenc}
\usepackage[T1]{fontenc}
\usepackage{mathptmx}
\usepackage{etoolbox}

\usepackage{mathtools}
\usepackage{amsthm}
\usepackage{graphicx}
\usepackage[dvipsnames]{xcolor}
\usepackage{enumitem}
\usepackage{microtype}
\usepackage[
    colorlinks=true,
    allcolors=blue
]{hyperref}

\makeatletter
\def\@email#1#2{%
 \endgroup
 \patchcmd{\titleblock@produce}
  {\frontmatter@RRAPformat}
  {\frontmatter@RRAPformat{\produce@RRAP{*#1\href{mailto:#2}{#2}}}\frontmatter@RRAPformat}
  {}{}
}%
\makeatother

\newtheorem{definition}{Definition}[section]
\newtheorem{lemma}{Lemma}[section]
\newtheorem{theorem}{Theorem}[section]
\newtheorem{corollary}{Corollary}[section]
\newtheorem{remark}{Remark}[section]
\newtheorem{proposition}{Proposition}[section]

\robustify{\sum}
\robustify{\prod}
\makeatletter
\apptocmd{\BibitemOpen}{%
  \edef\@currentlabel{\arabic{NAT@ctr}}%
  \edef\@currentHref{bibentry.\arabic{NAT@ctr}}%
  \Hy@raisedlink{\hyper@anchorstart{\@currentHref}\hyper@anchorend}%
}{}{\PackageWarning{venues/arxiv}{Could not patch \string\BibitemOpen}}
\makeatother

\newcommand{\KL}[2]{D_{\textsc{kl}}\left[#1 \parallel #2\right]}

\newcommand{\papertitle}{%
    Bayesian Replicator Dynamics and the Inversion\\
    of Strict Dominance in Bimatrix Games%
}
\newcommand{\papershorttitle}{Bayesian Replicator Dynamics}

\newcommand{\authorRuizSerra}{Jaime Ruiz-Serra}
\newcommand{\emailRuizSerra}{jaime.ruizserra@sydney.edu.au}

\newcommand{\authorAy}{Nihat Ay}

\newcommand{\authorHarre}{Michael Harr\'e}

\newcommand{\affilSydneyUnit}{Centre for Complex Systems}
\newcommand{\affilSydneyInst}{The University of Sydney}
\newcommand{\affilSydneyPlace}{Sydney, Australia}

\newcommand{\affilTUHHUnit}{Institute for Data Science Foundations}
\newcommand{\affilTUHHInst}{Hamburg University of Technology}
\newcommand{\affilTUHHPlace}{Hamburg, Germany}

\hypersetup{
    pdftitle={\papertitle},
    pdfauthor={\authorRuizSerra, \authorAy, \authorHarre},
}

\begin{document}

\title[\papershorttitle]{\papertitle}

\author{\authorRuizSerra}
\email[Author to whom correspondence should be addressed: ]{\emailRuizSerra}
\affiliation{\affilSydneyUnit, \affilSydneyInst, \affilSydneyPlace}

\author{\authorAy}
\affiliation{\affilTUHHUnit, \affilTUHHInst, \affilTUHHPlace}

\author{\authorHarre}
\affiliation{\affilSydneyUnit, \affilSydneyInst, \affilSydneyPlace}

\date{October 21, 2026}

\begin{abstract}
Replicator dynamics, the central model of evolutionary game theory, describe populations evolving under selection and agents learning by trial and error. They also admit a third use: continuous-time Bayesian inference, in which an agent planning against an opponent accumulates evidence for each candidate strategy. Here we show that this use divides all two-player games into two classes. With the standard \textit{expected-utility payoff}, the dynamics can be expressed as Bayesian updating only in games (with any number of actions) where no player has a dominant strategy. The \textit{logarithmic payoff} of Ay and Erb (2005) admits the Bayesian representation in every game. 
Coordination, anti-coordination, and cyclic games share qualitative dynamics under both payoff definitions, with equilibria displaced and selection thresholds shifted. In contrast, the flows of every game with a dominant strategy are reversed between the payoff definitions, with the strictly dominant strategy eliminated under logarithmic payoffs. Cooperation therefore emerges in the Prisoner's Dilemma, and games of common interest converge to mutual defection. 
For two-action games the classification is complete and rests on an exact Hamiltonian structure shared by both payoff definitions.
The structure gives a behavioural test: cycles in the classic Matching Pennies game have amplitude-independent period (isochronous) under logarithmic payoffs, but not under expected utility. Logarithmic evidence integration thus determines which strategic outcomes dynamical selection can reach, wherever learning approximates Bayesian updating.
 \end{abstract}

\maketitle

\section{Introduction}

The replicator equation is the standard model of frequency evolution under selection, and its state vector admits three interpretations: the composition of a biological population, the mix of strategies in play, or an agent's beliefs over competing hypotheses. In its canonical form the equation scores each type by an expected payoff. Scored instead by accumulated log-evidence, as a Bayesian reasoner scores hypotheses, it keeps the qualitative dynamics of coordination, anti-coordination, and cyclic games, but reverses the flow of every game with a dominant strategy. Cooperation then emerges in the Prisoner's Dilemma, and the Harmony game, in which cooperation is dominant, collapses to mutual defection. This paper proves where the inversion occurs and why. The rest of this section sets out the general replicator dynamics and their reading as continuous-time Bayesian inference (\S\ref{sec:bayesian-interpretation}). Section~\ref{sec:fitness-models-outlined} compares the expected-payoff (linear) and log-evidence (log-linear) fitness functions, characterises when the linear model can be expressed as Bayesian updating, and proves the inversion of pure-strategy dominance in every $n \times m$ game. Section~\ref{sec:analysis-2x2-games} classifies two-action games by whether the two models give topologically equivalent flows, and derives the Hamiltonian structure they share.

Formally, consider a system of $N$ interacting agents. The state of the $k$-th agent is a probability distribution $x^{(k)}$ over $n_k$ strategies, a point of the simplex $\mathcal{S}_{n_k} = \left\{ x \in \mathbb{R}^{n_k}_{\ge 0} \mid \sum_{i=1}^{n_k} x_i = 1 \right\}$. We write $\mathcal{S}^\circ_n$ for the relative interior of $\mathcal{S}_n$ (the open simplex, where all $x_i > 0$).

The evolution of the probability $x_i^{(k)}$ of strategy $i$ for agent $k$ is driven by its relative evolutionary success, or \textit{fitness}, compared to the population average. 
Let $f_i^{(k)}(\mathbf{x})$ denote the fitness of strategy $i$ for agent $k$, which depends on the state of all interacting agents, $\mathbf{x} = (x^{(1)}, \dots, x^{(N)})$. The general replicator dynamics are given by \cite{Hofbauer1998EvolutionaryGames}:
\begin{equation}\label{eq:general-replicator}
    \dot{x}_i^{(k)} = x_i^{(k)} \left( f_i^{(k)}(\mathbf{x}) - \bar{f}^{(k)}(\mathbf{x}) \right)
\end{equation}
where $\bar{f}^{(k)}(\mathbf{x}) = \sum_{j=1}^{n_k} x_j^{(k)} f_j^{(k)}(\mathbf{x})$ is the mean fitness of the agent's current distribution.

Focus on a single agent $k$ and drop the superscript for simplicity. This equation has two properties. First, the simplex is an invariant set under the dynamics: since $\sum \dot{x}_i = \sum x_i f_i(\mathbf{x}) - (\sum x_i) \bar{f}(\mathbf{x}) = \bar{f}(\mathbf{x}) - \bar{f}(\mathbf{x}) = 0$, the probabilities always sum to unity. Second, the growth rate of a strategy is determined by its excess fitness; strategies performing above average proliferate, while those performing below average decline.

\subsection{Bayesian Interpretation}\label{sec:bayesian-interpretation}

The equation began in population biology, tracking allele frequencies and behavioural phenotypes across generations \cite{Schuster1983ReplicatorDynamics, Taylor1978EvolutionaryStable, Zeeman1980PopulationDynamics}, and extends to coupled multi-population systems describing co-evolving species \cite{Hofbauer1998EvolutionaryGames, Hofbauer2003EvolutionaryGame} and multi-role asymmetric contests \cite{Taylor1979EvolutionarilyStable, Fudenberg1998TheoryLearning, Cressman2003EvolutionaryDynamics, Cressman2014ReplicatorEquation}. \citeauthor{Borgers1997LearningReinforcement}~\cite{Borgers1997LearningReinforcement} and \citeauthor{Sato2003CoupledReplicator}~\cite{Sato2003CoupledReplicator} shifted the focus from inheritance to reinforcement learning, showing that the same coupled dynamics describe the trial-and-error learning of independent agents \cite{Sato2002ChaosLearning}.

Dividing Equation~\eqref{eq:general-replicator} by $x_i$ shows that the rate of change of the agent's log-probability equals the excess fitness:
\begin{equation}
    \frac{d}{dt} \ln x_i 
    = \frac{1}{x_i} \frac{d}{dt} x_i 
    = f_i(\mathbf{x}) - \bar{f}(\mathbf{x})
\end{equation}
This elementary rearrangement carries the three structural facts used throughout the paper.
First, it shows the replicator equation to be a Bayesian filter\footnote{Bayesian filtering consists of the recursive application of Bayes' theorem to infer the hidden state of a system from noisy observations, updating the posterior distribution over time as observations are received.}, a view due to Shalizi~\cite{Shalizi2009DynamicsBayesian}, generalised to continuous-time information geometry by Harper~\cite{Harper2009InformationGeometry, Harper2010ReplicatorEquation} (see also \citeauthor{Czegel2022BayesDarwin}~\cite{Czegel2022BayesDarwin}). Under this interpretation, the strategy distribution $x \in \mathcal{S}_n$ is the agent's prior over a set of hypotheses, and the excess fitness is the rate of log-evidence accumulation.
Second, it makes log-odds the natural coordinates: in them the constrained simplex becomes an unconstrained space (\S\ref{sec:dual-space-mapping}).
Third, it motivates a fitness logarithmic in probabilities \cite{Ay2005NotionLinear} (\S\ref{sec:log-linear-fitness}).

Bayesian reasoning has also entered evolutionary game theory as a property of the players. \citeauthor{Patra2024InferringCooperate}~\cite{Patra2024InferringCooperate} study players in the repeated Prisoner's Dilemma who infer the strategy of a reactive opponent (one that responds to the previous round) by Bayesian updating, and identify conditions under which these Bayesian strategies are evolutionarily stable. \citeauthor{Patra2025BayesianEcoevolutionary}~\cite{Patra2025BayesianEcoevolutionary} bring games of incomplete information (Bayesian games) into eco-evolutionary dynamics. In both, Bayesian reasoning is part of the players' strategies or of the game, and selection acts on it in the usual way. In the present model the replicator equation itself performs the Bayesian update.

\subsection{Strategic Planning and Opponent Modelling}

With the Bayesian-filter interpretation in place, consider an agent, Agent A, planning its action against another agent, Agent B. No actions are exchanged: the coupled system is a planning process internal to Agent A, its reasoning about the interaction \textit{before} any action is selected.
We extend the single-agent replicator-as-inference equivalence to a closed-loop planning setting, in which the simulated action choices of A and B update one another.

Agent A holds a belief over joint strategies\footnote{Hereafter, we will stop using $\mathbf{x}$ in favour of using $x$ and $y$ explicitly, since we focus on the two-agent case.} $\mathbf{x} = (x, y)$, its intended strategy $x \in \mathcal{S}_n$ and its prediction $y \in \mathcal{S}_m$ of Agent B's strategy.
The fitness $f_i : \mathcal{S}_n \times \mathcal{S}_m \to \mathbb{R}$ is the \textit{anticipated} utility, or simulated evidence, supporting action $i$ given the strategies $(x, y)$. Strategies are reinforced by their performance (utility if chosen) in Agent A's internal simulation, turning the `Imitation of Success' protocol of population games~\cite{Sandholm2010PopulationGames} into intra-agent Bayesian belief updating without observations.

Agent A represents the rates of change of both strategies by the coupled replicator dynamics
\begin{equation}
\begin{aligned}\label{eq:general-coupled-replicator}
    \dot{x}_i &= x_i \left( f_i^A(x, y) - \bar{f}^A(x, y) \right) \\
    \dot{y}_j &= y_j \left( f_j^B(x, y) - \bar{f}^B(x, y) \right)
\end{aligned}
\end{equation}
The simulated strategy $x$ enters the fitness $f_j^B(x, y)$ that updates the belief $y$, and $y$ enters the fitness $f_i^A(x, y)$ that updates $x$, closing the loop.
The functional form of the fitness $f_i$ determines the dynamical behaviour of the system \eqref{eq:general-coupled-replicator}. In the following section we introduce the two fitness forms this paper is concerned with.

Throughout the paper, the dynamics assume infinite precision (no noise or bounded rationality), and each belief stream integrates evidence from a non-stationary counterpart without temporal discounting. We return to both limitations in the concluding section.

\section{Functional Form of Fitness}\label{sec:fitness-models-outlined}

We contrast two models:
\begin{enumerate}
\item \textbf{Linear Model (\S\ref{sec:linear-fitness}):} the canonical case (Eq.~\ref{eq:fitness-euclidean-gradient}) with fitness linear in the state probabilities. This is the standard expected-utility formulation.
\item \textbf{Log-Linear Model (\S\ref{sec:log-linear-fitness}):} an information-theoretic model \cite{Ay2005NotionLinear} (Eq.~\ref{eqn:log-linear-fitness}), with fitness logarithmic in probabilities\footnote{In \citeauthor{Ay2005NotionLinear}~\cite{Ay2005NotionLinear} these dynamics are termed ``linear replicator equations'', because the fitness is affine in the exponential-affine (logarithmic) coordinates. We say ``log-linear'' throughout, since the fitness is linear in log probabilities, to distinguish this model from the classical linear expected-utility formulation.}.
\end{enumerate}

Both models employ payoff matrices, e.g. $A \in \mathbb{R}^{n\times m}$ for Agent A, and both drive the same update rule \eqref{eq:general-coupled-replicator}. Only the fitness $f_i$ changes, and with it the evidence the update integrates. Informally, the log-linear fitness treats each strategy as a hypothesis about the opponent and scores it by the log-likelihood of the behaviour it predicts, whilst the linear fitness scores each strategy by its expected payoff, which serves as such a log-likelihood only in some games. \S\ref{sec:linear-fitness} and \S\ref{sec:log-linear-fitness} introduce the two models, and \S\ref{sec:gauge} the payoff transformations that leave the dynamics unchanged; \S\ref{sec:bayesian-evidence} then develops the Bayesian interpretation of both and identifies the games in which the linear model admits it.

\subsection{The Linear Model}\label{sec:linear-fitness}

In the classical case, agents value strategies by expected utility. The objective is $F^A(x, y) = x^{\top} A y$, and the fitness of strategy $i$ is its expected payoff against the opponent's current mixture,
\begin{equation}\label{eq:fitness-euclidean-gradient}
    f^A_i(x, y) = (A y)_i .
\end{equation}
Agent A models Agent B as having its own payoff matrix $B \in \mathbb{R}^{m \times n}$, written from B's perspective, and maximising $F^B(x, y) = y^{\top} B x$. With these fitnesses, the coupled dynamics \eqref{eq:general-coupled-replicator} are the classic bimatrix replicator equations \cite{Hofbauer1998EvolutionaryGames}:
$$\begin{aligned} \dot{x}_i &= x_i \left( (Ay)_i - x^{\top} A y \right) \\ \dot{y}_j &= y_j \left( (Bx)_j - y^{\top} B x \right). \end{aligned}$$

\subsection{The Log-Linear Model}\label{sec:log-linear-fitness}

Following \citeauthor{Ay2005NotionLinear}~\cite{Ay2005NotionLinear}, the log-linear model replaces the expected payoff with a fitness linear in the \emph{log}-probabilities of the opponent's actions. The fitness of action $i$ is
\begin{equation}\label{eqn:log-linear-fitness}
    f^A_i(x, y) = (A \ln y)_i = \sum_{k=1}^m A_{ik} \ln y_k .
\end{equation}
The update rule remains the replicator equation, changing only the fitness function, so the dynamics for Agent A are
\begin{equation}
    \dot{x}_i = x_i \Bigg( \underbrace{(A \ln y)_i}_{\text{Perceived Value}} - \underbrace{\sum_{k=1}^n x_k (A \ln y)_k}_{\text{Expected Baseline}} \Bigg)
\end{equation}
(and likewise for B).

The logarithm diverges as probabilities approach absolute certainty, imposing a state-dependent penalty there. This conditionally inverts or accelerates the dynamics at the boundary, altering the global topology of the phase space in some cases, which are identified in Theorem~\ref{thm:partition}. Proposition~\ref{prop:compactification} shows, for the two-action case of \S\ref{sec:analysis-2x2-games}, that the diverging trajectories map continuously to boundary equilibria. The dynamics have well-defined signed limits at the simplex boundaries, requiring no artificial smoothing or regularisation to evaluate the system.

\subsection{Payoff Gauge}\label{sec:gauge}

Not all of the payoff matrix affects the dynamics. The replicator equations are invariant under the transformations
\begin{equation*}
    A \mapsto A + \mathbf{1}_n c^{\top}, \qquad B \mapsto B + \mathbf{1}_m d^{\top}, \qquad c \in \mathbb{R}^m, \; d \in \mathbb{R}^n,
\end{equation*}
because each adds a common term to every component of the corresponding fitness vector, and the common term cancels in the excess fitness \cite{Hofbauer1998EvolutionaryGames}. 

The invariance goes beyond uniform payoff shifts. Under the log-linear model, the added common term is $c^{\top} \ln y$. We call these \emph{gauge transformations}, borrowing the physics term for changes of parameters that do not alter a model's observables, and call payoff matrices related by them \emph{gauge equivalent}. The term carries no further technical content and every claim below about representing a payoff matrix is a claim modulo this gauge. \citeauthor{Weibull1997EvolutionaryGame}~\cite[\S1.3.3]{Weibull1997EvolutionaryGame} calls the same transformations \emph{local shifts} of payoffs, local meaning specific to each opponent strategy.

\subsection{Fitness as Bayesian Evidence}\label{sec:bayesian-evidence}

The excess fitness is the rate at which the agent accumulates log-evidence (\S\ref{sec:bayesian-interpretation}), and the log-linear fitness makes that evidence Bayesian. Throughout the paper, the agent's own mixed strategy $x$ is the prior over hypotheses (one per available action), the opponent's mixture $y$ generates the evidence, and the payoff row $A_{i \cdot}$ (a composition, a vector of possibly fractional pseudo-counts of opponent actions) is the data that hypothesis $i$ predicts. Framed this way, the log-linear fitness consists of scoring strategy $i$ by the log-likelihood of the composition $A_{i \cdot}$ under $y$,
\begin{equation*}
    f^A_i(x, y) = \sum_{k=1}^m A_{ik} \ln y_k = \ln \prod_{k=1}^m y_k^{A_{ik}} = \ln l_i(y),
\end{equation*}
where $l_i(y)$ is the multinomial likelihood\footnote{Here $l_i$ is the likelihood of an ordered i.i.d.\ sample with composition $A_{i\cdot}$, equivalently the kernel of the multinomial counts likelihood without its combinatorial coefficient; Remark~\ref{rem:multinomial-coefficient} discusses restoring that coefficient.} in which the payoff row $A_{i \cdot}$ acts as the count vector, making strategy $i$ the hypothesis that the agent will encounter $A_{ik}$ occurrences (which may be fractional, but negative entries are removed by the gauge, \S\ref{sec:gauge}) of opponent action $k$. The log-linear replicator is then the continuous-time limit of Bayesian updating over the strategy set with likelihoods $l_i$, in the sense of the replicator-as-Bayesian-filter interpretation of \S\ref{sec:bayesian-interpretation}.

Under the likelihood interpretation, the gauge multiplies every $l_i$ by the common factor $\prod_k y_k^{c_k}$, which cancels in the Bayesian posterior. The operative content of the fitness is the likelihood \emph{ratios} $l_i / l_j$, and these are gauge-invariant. Negative payoff entries are removed by the same mechanism, as choosing $c_k \ge \max_i(-A_{ik})$ gives a gauge-equivalent matrix with every entry non-negative. Each bimatrix game, including the zero-sum games of \S\ref{sec:analysis-2x2-games}, has a representative in which every row is a pseudo-count vector.

The rows of $A$ cannot, by contrast, be normalised into probability distributions without altering the dynamics, since rescaling a single row ($f_i \mapsto \lambda_i f_i$) changes relative fitness differences. Any interpretation that treats an individual row as a normalised target distribution therefore depends on the arbitrary gauge, whereas likelihood ratios do not.\footnote{For instance, in a Prisoner's Dilemma with payoffs $(R, S, T, P) = (3, 0, 5, 1)$, row-normalising gives the profiles $(1, 0)$ and $(5/6, 1/6)$, whilst first shifting all payoffs by $+10$ (with $c = (10, 10)$ and $A \mapsto A + \mathbf{1} c^\top$) gives $(13/23, 10/23)$ and $(15/26, 11/26)$. The dynamics for these two examples are identical (same orbits, velocities, etc.), but their profiles are different (e.g. $(1,0) \ne (13/23, 10/23)$).}

Under the same interpretation, the canonical linear model has a mismatch with evidence integration. The fitness $f_i(\mathbf{x})$ drives a log-probability update ($\frac{d}{dt} \ln x_i$), so exponentiating it recovers the likelihood the dynamics attribute to strategy $i$. For the linear model this is $\exp\big((Ay)_i\big)$, an exponentiated mixture of the opponent's raw probabilities, which is not the likelihood kernel of a sampling model over opponent actions, whatever pseudo-counts are chosen.

Transposing the terms rehabilitates the linear model, conditionally. Framing it as integrating evidence \textit{about} the opponent, each hypothesis $i$ carries its own predictive distribution $p_i(\cdot)$ over opponent actions, and setting $A_{ik} = \ln p_i(k)$ makes $(Ay)_i = \sum_k y_k \ln p_i(k)$ an expected log-likelihood. This is the replicator-as-inference of \citeauthor{Shalizi2009DynamicsBayesian} and \citeauthor{Harper2009InformationGeometry} \cite{Shalizi2009DynamicsBayesian, Harper2009InformationGeometry}, which we call the \emph{conventional} Bayesian representation. The substantive question is \emph{whether a given payoff matrix accepts the conventional Bayesian interpretation, allowing for the transformations under which the dynamics are invariant} (\S\ref{sec:gauge}). We show that the answer is determined by payoff dominance: between pure strategies for a two-action agent, and between mixed strategies of the exponentiated payoffs $\exp(A)$ in general (Proposition~\ref{prop:representability-nxm}).

\begin{proposition}[Dominance obstructs the conventional Bayesian representation]\label{prop:conventional-obstruction}
Let $A \in \mathbb{R}^{n \times m}$ and suppose row $i$ dominates row $j$ entrywise: $A_{ik} \ge A_{jk}$ for all $k$, with at least one strict inequality. Then no gauge representative $A + \mathbf{1}c^{\top}$, $c \in \mathbb{R}^m$, has every row equal to the entrywise logarithm of a probability vector; equivalently, no assignment of predictive distributions renders the linear fitness, modulo the gauge, the expected log-likelihood $\sum_k y_k \ln p_i(k)$ for every strategy simultaneously.
\end{proposition}

\begin{proof}
Such a representative requires $u \in \mathbb{R}^m_{>0}$, with $u_k = \exp(c_k)$, satisfying $\sum_k \exp(A_{ik}) u_k = 1$ for every row. Entrywise dominance gives $\sum_k \big(\exp(A_{ik}) - \exp(A_{jk})\big) u_k > 0$, every summand being non-negative and at least one strictly positive, so the two row sums cannot both equal one.
\end{proof}

In \S\ref{sec:analysis-2x2-games} we characterise a partition of two-action games based on this obstruction: the conventional representation exists if and only if the agent has no weakly dominant pure strategy (Proposition~\ref{prop:representability}). The linear model is therefore a Bayesian filter only on part of the payoff space, and its boundary also delineates whether the dynamics under the linear and log-linear fitness models are topologically equivalent or inverse.

Beyond two actions, comparing pure strategies does not settle the question. The criterion compares mixed strategies in the exponentiated payoffs $M = \exp(A)$, taken entrywise, rather than in $A$. Say that a mixed strategy $p \in \mathcal{S}_n$ \emph{weakly dominates} $q \in \mathcal{S}_n$ in a matrix $M \in \mathbb{R}^{n \times m}$ if $p^{\top} M \ge q^{\top} M$ entrywise, with at least one strict inequality.

\begin{proposition}[Conventional representability in $n \times m$ games]\label{prop:representability-nxm}
Let $A \in \mathbb{R}^{n \times m}$ and $M = \exp(A)$ entrywise. The following are equivalent:
\begin{enumerate}
    \item[(i)] some gauge representative $A + \mathbf{1}c^{\top}$, $c \in \mathbb{R}^m$, has every row equal to the entrywise logarithm of a probability vector;
    \item[(ii)] some completely mixed opponent strategy $\beta \in \mathcal{S}^\circ_m$ equalises the exponentiated payoffs, $(M\beta)_i = (M\beta)_j$ for all $i, j$;
    \item[(iii)] no mixed strategy of the agent weakly dominates another in $M$.
\end{enumerate}
\end{proposition}

\begin{proof}
(i)$\Leftrightarrow$(ii). As in the proof of Proposition~\ref{prop:conventional-obstruction}, (i) holds if and only if some $u \in \mathbb{R}^m_{>0}$ solves $Mu = \mathbf{1}$, with $u_k = \exp(c_k)$. Such a $u$ gives $\beta = u / \sum_k u_k$, with common value $(M\beta)_i = 1 / \sum_k u_k$. Conversely, if $M\beta = t\mathbf{1}$ with $\beta \in \mathcal{S}^\circ_m$, then $t > 0$ because $M$ is entrywise positive, and $u = \beta / t$ solves $Mu = \mathbf{1}$.

(ii)$\Leftrightarrow$(iii). Condition (ii) asks for a vector $z = (\beta, t) \in \mathbb{R}^{m+1}$, with every entry strictly positive, in the kernel of the $n \times (m+1)$ matrix $[M, -\mathbf{1}]$. By Stiemke's lemma \cite{Stiemke1915UeberPositive}, for any real matrix $K$ exactly one of two alternatives holds: $Kz = 0$ for some $z$ with every entry strictly positive, or $\lambda^{\top} K$ is entrywise non-negative and non-zero for some vector $\lambda$. For $K = [M, -\mathbf{1}]$ the second alternative reads $\lambda^{\top} M \ge 0$ entrywise and $\sum_i \lambda_i \le 0$, with not all of these $m + 1$ quantities zero. Suppose it holds, and split $\lambda = \lambda^+ - \lambda^-$ into non-negative parts with disjoint supports. Then $\lambda^- \ne 0$, since otherwise $\sum_i \lambda_i \le 0$ forces $\lambda = 0$; and $\lambda^+ \ne 0$, since otherwise every entry of $\lambda^{\top} M = -(\lambda^-)^{\top} M$ is strictly negative. With $a = \sum_i \lambda^+_i$ and $b = \sum_i \lambda^-_i$, so that $0 < a \le b$, the mixed strategies $p = \lambda^+ / a$ and $q = \lambda^- / b$ satisfy
\begin{equation*}
    p^{\top} M \;\ge\; \tfrac{b}{a}\, q^{\top} M \;\ge\; q^{\top} M .
\end{equation*}
If $a < b$, the second inequality is strict in every entry, because $q^{\top} M$ is entrywise positive. If $a = b$, the first is strict in some entry, because $\lambda^{\top} M \ne 0$. Either way $p$ weakly dominates $q$ in $M$, so (iii) fails. Conversely, if $p$ weakly dominates $q$ in $M$, then $\lambda = p - q$ satisfies the second alternative with $\sum_i \lambda_i = 0$, so (ii) fails.
\end{proof}

Condition (ii) says that the representation exists if and only if some completely mixed opponent strategy leaves the agent indifferent among all of its pure strategies, with payoffs $\exp(A)$ in place of $A$. The equivalence of (ii) and (iii) is the full-support case of Weinstein's characterisation of best-reply sets \cite{Weinstein2020BestreplySets}, applied to the payoffs $M$ with every pure strategy in the set; the proof above keeps the argument self-contained. Proposition~\ref{prop:conventional-obstruction} is the case of pure $p$ and $q$: the exponential is strictly increasing, so it preserves entrywise dominance between two rows. Its convexity extends the obstruction to a pure strategy $j$ dominated by a mixture in $A$. If $p^{\top} A \ge A_{j\cdot}$ entrywise with a strict entry, Jensen's inequality gives $p^{\top} M \ge \exp(p^{\top} A) \ge M_{j\cdot}$ entrywise, with the strict entry preserved, so (iii) fails. The same convexity lets a mixture dominate in $M$ without dominating in $A$, which is why the criterion must be stated in $\exp(A)$ (Remark~\ref{rem:no-dominance-insufficient}).

For a two-action agent the criterion reduces to pure strategies, whatever the number of opponent actions.

\begin{corollary}[Two-action agents]\label{cor:representability-2xm}
Let $n = 2$ and $m \ge 2$. The conventional representation exists if and only if neither pure strategy weakly dominates the other in $A$, that is, if and only if the row difference $A_{1\cdot} - A_{2\cdot}$ is zero or has entries of both strict signs.
\end{corollary}

\begin{proof}
For $n = 2$, any $p \ne q$ in $\mathcal{S}_2$ give $p - q = (p_1 - q_1)(1, -1)$ with $p_1 \ne q_1$, so a mixed strategy weakly dominates another in $M$ if and only if one row of $M$ weakly dominates the other. The exponential is strictly increasing, so this happens if and only if one row of $A$ weakly dominates the other, and Proposition~\ref{prop:representability-nxm} applies.
\end{proof}

\begin{remark}[Absence of dominance does not suffice beyond two actions]\label{rem:no-dominance-insufficient}
For $n \ge 3$ an agent can have no dominated strategy of any kind and still lack a conventional representation. For $n > m$ this happens for a trivial reason. Representability puts $\mathbf{1}$ in the column space of $M$, so the $n \times (m+1)$ matrix $[M, \mathbf{1}]$ has rank at most $m$. For $n > m$ this is a non-trivial polynomial condition on the entries of $M$, so it holds only on a set of Lebesgue measure zero (a set that randomly drawn payoffs fall in with probability zero), and $A \mapsto \exp(A)$ is a diffeomorphism (a smooth one-to-one map whose inverse is also smooth) onto the entrywise positive matrices. Almost every game in which the agent has more strategies than its opponent is therefore non-representable, whether or not it has a dominated strategy.

Square games are not exempt. Take
\begin{equation*}
    A = \begin{pmatrix} 0 & 2 & 1 \\ 2 & 0 & 1 \\ 1 & 1 & 1 \end{pmatrix}.
\end{equation*}
Against the uniform opponent strategy every row of $A$ earns expected payoff $1$. No mixed strategy therefore weakly dominates another in $A$: averaging a dominance $p^{\top} A \ge q^{\top} A$ with a strict entry over the uniform strategy would give $1 > 1$. In $M = \exp(A)$, however, the even mixture of the first two rows earns $\big((1 + e^2)/2,\, (1 + e^2)/2,\, e\big)$, which exceeds the third row $(e, e, e)$ by $(e - 1)^2/2 \approx 1.48$ in the first two entries and equals it in the third. By Proposition~\ref{prop:representability-nxm}, no conventional representation exists. The direct computation agrees: $Mu = \mathbf{1}$ forces $u_1 = u_2$ and then $u_1 (e - 1)^2 = 0$, so the unique solution $u = (0, 0, e^{-1})$ puts all of its weight on one opponent action, which no finite gauge vector $c$ attains. The obstruction is the convexity gap between mixing rows before and after exponentiating.
\end{remark}

\subsection{Dominance Inversion}

We close this section with a structural consequence of the fitness model itself, valid in every $n \times m$ game. Under linear fitness, entrywise dominance of one payoff row over another makes the dominating strategy the better reply against every opponent behaviour. Under log-linear fitness this ordering is reversed: everywhere on the interior, and in every bimatrix game.

\begin{proposition}[Inversion of pure-strategy dominance]\label{prop:dominance-inversion}
Let $A \in \mathbb{R}^{n \times m}$ and suppose row $i$ dominates row $j$ entrywise: $A_{ik} \ge A_{jk}$ for all $k$, with at least one strict inequality. Then, under the log-linear fitness \eqref{eqn:log-linear-fitness},
$$
f^A_i(x, y) - f^A_j(x, y) = \sum_{k=1}^m \big(A_{ik} - A_{jk}\big) \ln y_k < 0
$$
for every $y \in \mathcal{S}^\circ_m$, so that $\tfrac{d}{dt} \ln(x_i / x_j) < 0$ along every interior trajectory: the dominant strategy loses probability share to the dominated one throughout the interior.
\end{proposition}

\begin{proof}
Every coefficient $A_{ik} - A_{jk}$ is non-negative and at least one is strictly positive, whilst $\ln y_k < 0$ on the open simplex, so the sum is strictly negative. The replicator equation gives $\tfrac{d}{dt} \ln(x_i / x_j) = f^A_i - f^A_j$.
\end{proof}

\begin{corollary}[Elimination of strictly dominant strategies]\label{cor:strict-elimination}
Suppose strategy $i$ strictly dominates every other strategy of Agent A: $A_{ik} > A_{jk}$ for all $j \ne i$ and all $k$, with $m \ge 2$. Then, under the log-linear fitness \eqref{eqn:log-linear-fitness}, $x_i(t) \to 0$ along every interior trajectory, so the strictly dominant strategy is eliminated.
\end{corollary}

\begin{proof}
Fix $j \ne i$ and set $\delta_j = \min_k (A_{ik} - A_{jk}) > 0$. Each coefficient in the sum of Proposition~\ref{prop:dominance-inversion} is at least $\delta_j$ and each $\ln y_k$ is negative, so
$$f^A_i - f^A_j \;\le\; \delta_j \sum_{k=1}^m \ln y_k \;\le\; -\,\delta_j\, m \ln m,$$
the second inequality because $\prod_k y_k \le m^{-m}$ on the simplex. Hence $\ln(x_i/x_j)$ decreases at least linearly in $t$, so $x_i/x_j \to 0$ for every $j \ne i$, and $x_i = \big(1 + \sum_{j \ne i} x_j/x_i\big)^{-1} \to 0$.\footnote{Each log-coordinate velocity is bounded by an affine function of the largest absolute log-probability, so trajectories stay in the interior of the simplex in finite time.}
\end{proof}

The uniform rate only applies to strict dominance. If the inequality $A_{ik} \ge A_{jk}$ is strict only in some columns, the fitness gap can vanish as the opponent concentrates on the others, and the ratio $x_i/x_j$ may converge to a positive limit. Elimination cannot be extended to weakly dominant strategies (see two-action stalling orbits in Appendix~\ref{apx:partition-proof}).

Entrywise row dominance is dominance against every pure opponent strategy and hence, the payoff comparison being linear in $y$ under the linear model, against every mixed strategy as well. Thus logarithmic evidence integration reverses pure-strategy dominance in every $n \times m$ bimatrix game. The result requires no special coordinates and no restriction to two actions. It can be explained informationally as follows. A payoff row with larger entries is a hypothesis that predicts more, which is costlier when the predicted events carry small probability. Theorem~\ref{thm:partition} below is the two-action case of this general fact, locating where the reversal reshapes the global topology of the joint dynamics (Set $\mathcal{D}$) and where an interior equilibrium absorbs it (Set $\mathcal{E}$).

\begin{remark}[Restoring the multinomial coefficient]\label{rem:multinomial-coefficient}
The fitness \eqref{eqn:log-linear-fitness} omits the combinatorial coefficient of the multinomial counts likelihood. Restoring it would add the constant $\ln (N_i! / \prod_k A_{ik}!)$, with $N_i = \sum_k A_{ik}$, to the fitness of strategy $i$. Such a constant leaves the boundary orientations intact but moves interior nullclines. When one strategy dominates another, it can do more. Under \eqref{eqn:log-linear-fitness}, the dominated strategy has the higher fitness at every interior state (Proposition~\ref{prop:dominance-inversion}), but its lead varies from state to state. If the coefficient favours the dominant strategy by more than this lead at some state, the dominant strategy is ahead again near that state, and the inversion fails there. In a two-action game, this creates new interior nullclines. The coefficient also depends on the gauge (\S\ref{sec:gauge}). Raising every payoff by the same amount leaves the dynamics under \eqref{eqn:log-linear-fitness} unchanged, but it changes the coefficient. Whether the inversion fails can therefore depend on an arbitrary payoff shift.

The Prisoner's Dilemma with $(R, S, T, P) = (3, 0, 5, 1)$ shows both effects. Under \eqref{eqn:log-linear-fitness}, cooperation's lead over defection is smallest, about $1.91$, when the opponent cooperates with probability $y_1 = 2/3$. At these payoffs, the coefficient favours defection by $\ln 6 \approx 1.79$. This is less than $1.91$, so the inversion survives everywhere. If every payoff is first raised by $20$, the coefficient favours defection by about $1.98$. This is more than $1.91$, so each agent gains two interior nullclines, and the symmetric game gains four interior rest points: two saddles and two centres. The results of this paper concern the fitness \eqref{eqn:log-linear-fitness} as stated.
\end{remark}

Proposition~\ref{prop:dominance-inversion} resembles, in its conclusion but not its mechanism, an established literature on the failure of dominance elimination. Under linear fitness, imitative dynamics of replicator form eliminate strictly dominated pure strategies \cite{SamuelsonZhang1992EvolutionaryStability}, but there are two known ways that overturn the elimination.

The first is stochastic. In the aggregate-shock formulation of \citeauthor{FudenbergHarris1992AggregateShocks}~\cite{FudenbergHarris1992AggregateShocks}, an It\^o correction penalises growth rates by the variance, so ranking by expected payoff need not survive, though dominated strategies are still eliminated under suitable conditions \cite{Imhof2005LongrunBehavior, HofbauerImhof2009TimeAverages}. \citeauthor{MertikopoulosViossat2016PayoffShocks}~\cite{MertikopoulosViossat2016PayoffShocks} show that under payoff shocks dominated strategies survive in the high-noise regime but become extinct when the noise is low, or at any noise level if agents revise on cumulative rather than instantaneous payoffs. The second route is protocol-based: \citeauthor{Mertikopoulos2022SurvivalDominated}~\cite{Mertikopoulos2022SurvivalDominated} show that imitation protocols favouring rare strategies sustain strictly dominated strategies at non-trivial frequencies. Both routes can do more than let a dominated strategy survive. In the Prisoner's Dilemma with aggregate shocks, \citeauthor{Imhof2005LongrunBehavior}~\cite{Imhof2005LongrunBehavior} shows that defection becomes extinct and cooperation fixes almost surely once the noise on defection exceeds that on cooperation by enough, namely $\tfrac12(\sigma_D^2-\sigma_C^2) > \max\{T - R,\, P - S\}$, the larger of the two boundary incentives $\Delta_0$, $\Delta_1$ of \S\ref{sec:topological-equivalence}; and under imitation protocols that favour frequent strategies a dominated strategy can take over from an initial majority, provided its payoff disadvantage is small \cite{Mertikopoulos2022SurvivalDominated}. In each case the outcome is conditional on a noise asymmetry or on the revision protocol.

Here the mechanism is different. The revision protocol remains the unmodified replicator and the dynamics are deterministic. The reversal needs no noise and no parameter: it holds everywhere on the interior, in every $n \times m$ game, and follows from the fitness function alone. There is no conflict with the elimination results: dominance in Proposition~\ref{prop:dominance-inversion} is a relation between rows of the payoff matrix $A$, whereas the flow is driven by the log-evidence score $A \ln y$, and the two orderings coincide only under linear fitness. Proposition~\ref{prop:dominance-inversion} states that they are opposed.

Ours is a modelling distinction rather than a normative claim: the linear model computes an expected payoff, the log-linear model accumulates log-evidence. Our results caution against treating the Bayesian variant as normatively superior, as the same mechanism that sustains cooperation in the Prisoner's Dilemma removes it in the Harmony game (\S\ref{sec:topological-equivalence}). 
\section{Classification of Two-Action Games}\label{sec:analysis-2x2-games}

We now restrict to $2 \times 2$ bimatrix games and compare the two flows. As a benchmark, recall the linear case: every Nash equilibrium is a rest point of the replicator dynamics, every strict Nash equilibrium is asymptotically stable, and an interior orbit can converge only to a Nash equilibrium (the Folk Theorem of evolutionary game theory \cite[Theorem 7.2.1]{Hofbauer1998EvolutionaryGames}). Every pure-strategy profile is also a rest point, so a strictly dominated strategy is a fixed point, though an unstable one \cite{Hofbauer1998EvolutionaryGames}.

We develop our analysis over three stages: \S\ref{sec:dual-space-mapping} sets up log-odds coordinates, \S\ref{sec:topological-equivalence} partitions $2\times 2$ games by whether the log-linear flow is homeomorphic to the linear one (related to it by a continuous deformation of the state space with a continuous inverse), and \S\ref{sec:interaction-curvature}--\ref{sec:hamiltonian-structure} reduce each game to two parameters and give the Hamiltonian that governs the whole class.

\subsection{Log-Odds Coordinates}\label{sec:dual-space-mapping}

Each agent's mixed strategy has one degree of freedom. We track Agent A by the log-odds $v = \ell(x) = \ln(x_1/x_2)$ and Agent B by $w = \ln(y_1/y_2)$; the inverse is the sigmoid, $x_1 = \sigma(v) = 1/(1 + e^{-v})$. These coordinates are unconstrained: the open simplex $\mathcal{S}_2^\circ$ maps onto the whole real line $\mathbb{R}$, and the boundary, where a probability is zero and the logarithm diverges, is pushed to infinity. Strategic dynamics often converge to pure strategies on that boundary, so we record how the map behaves there.

\begin{proposition}[Log-Odds Map Extended to Boundary Equilibria]\label{prop:compactification}
Let $\bar{\mathbb{R}} = [-\infty, +\infty]$ denote the real line composed with $\pm \infty$.
Let $\mathcal{S}^\circ_n = \{x \in \mathbb{R}_{>0}^n \mid \sum_{i=1}^n x_i = 1\}$ denote the open simplex.
The scalar log-odds map $\ell: \mathcal{S}^\circ_2 \to \mathbb{R}$, such that $v = \ell(x) = \ln(x_1/x_2)$, extends to a unique bijective map $\bar{\ell}: \mathcal{S}_2 \to \bar{\mathbb{R}}$ on the closed simplex,
$$
\bar{v} = \bar{\ell}(x) = \begin{cases}
-\infty & \text{if } x_1 = 0 \\
\ln(x_1/x_2) & \text{if } x_1 \in (0,1) \\
+\infty & \text{if } x_1 = 1
\end{cases}
$$
\end{proposition}

\begin{proof}
The inverse map $\ell^{-1}: \mathbb{R} \to \mathcal{S}^\circ_2$, whose first component is the sigmoid $x_1 = \sigma(v) = \frac{1}{1+e^{-v}}$, is continuous and strictly monotonic. Evaluating sequential limits at the infinity boundaries gives $\lim_{v \to -\infty} x_1 = 0$ and $\lim_{v \to +\infty} x_1 = 1$.
Extending this map by its boundary limits gives a continuous bijection $\bar{\ell}^{-1}: \bar{\mathbb{R}} \to \mathcal{S}_2$ from the compact space $\bar{\mathbb{R}} \cong [0,1]$ to the Hausdorff space $\mathcal{S}_2$; a continuous bijection from a compact space to a Hausdorff space is a homeomorphism, so the inverse map $\bar{\ell}$ is also continuous.
Any trajectory that diverges to infinity in these coordinates therefore maps to a boundary equilibrium (such as a pure strategy) on the edges of the probability simplex.
\end{proof}

\subsection{Topological Equivalence}\label{sec:topological-equivalence}

Let $\mathcal{M} = \mathcal{S}^\circ_n \times \mathcal{S}^\circ_m$ define the open joint strategy manifold, and let $\bar{\mathcal{M}}$ denote its closure. We compare the two flows on the interior $\mathcal{M}$, because the log-linear vector field is undefined wherever a coordinate vanishes. The behaviour at the boundary enters only through the limits of Proposition~\ref{prop:compactification}. Consider the continuous phase flows\footnote{
    The phase flow $\phi: \mathcal{M} \times \mathbb{R} \to \mathcal{M}$ represents the global integration of the vector field $\mathcal{X}$. The second argument is time: $\phi(z, t)$ is the position after time $t$ of the system started at state $z$. Both flows are complete (defined for all times) on the interior: in log-odds coordinates each velocity component grows at most linearly in the opponent's coordinate, precluding finite-time escape.
}$$
\phi_{\mathrm{Lin}}: \mathcal{M} \times \mathbb{R} \to \mathcal{M}
\quad \text{and} \quad
\phi_{\mathrm{Log}}: \mathcal{M} \times \mathbb{R} \to \mathcal{M},
$$ 
generated by the linear vector field $\mathcal{X}_{\mathrm{Lin}}$ and the log-linear vector field $\mathcal{X}_{\mathrm{Log}}$, respectively.

\begin{definition}[Topological equivalence]\label{def:topological-equivalence}
The dynamical system $(\mathcal{M}, \mathcal{X}_{\mathrm{Log}})$ is defined as \textit{topologically equivalent} to $(\mathcal{M}, \mathcal{X}_{\mathrm{Lin}})$ if there exists a homeomorphism $h : \mathcal{M} \to \mathcal{M}$ that maps the unparameterised trajectories\footnote{
    i.e., ordered sets of points $\mathcal{O} = \{p \in \mathcal{M}\}$, rather than parameterised functions $z(t)$
} of the log-linear flow bijectively onto the trajectories of the linear flow whilst preserving their directional orientation in time.

Formally, letting $\mathcal{O}(z, \phi) = \{\phi(z, t) \mid t \in \mathbb{R}\}$ denote the complete orbit through an initial joint belief state $z \in \mathcal{M}$, this mapping requires that
$$
    h\big(\mathcal{O}(z, \phi_{\mathrm{Log}})\big) = \mathcal{O}\big(h(z), \phi_{\mathrm{Lin}}\big) 
    \quad 
    \forall z \in \mathcal{M}.
$$
Furthermore, the homeomorphism must preserve the temporal orientation of the flow. Formally, there must exist a continuous function $\tau : \mathcal{M} \times \mathbb{R} \to \mathbb{R}$ such that, for each fixed $z \in \mathcal{M}$, the map $t \mapsto \tau(z, t)$ is strictly increasing with $\tau(z, 0) = 0$, and
$$
    h\!\left(\phi_{\mathrm{Log}}(z, t)\right) = \phi_{\mathrm{Lin}}\!\left(h(z),\, \tau(z, t)\right) 
    \quad 
    \forall z \in \mathcal{M},\, t \in \mathbb{R}.
$$
The strict monotonicity of $\tau(z, \cdot)$ guarantees that $h$ sends future states of the log-linear flow to future states of the linear flow, preserving the common arrow of time while permitting the two flows to traverse their respective orbits at different (point-dependent) speeds.

To preserve the identity of the strategies (a reflection must not equate an attractor at mutual cooperation with one at mutual defection), the homeomorphism $h$ is required to be \textit{vertex-preserving}: $h$ must admit a continuous extension $\bar{h}: \bar{\mathcal{M}} \to \bar{\mathcal{M}}$ satisfying $\bar{h}(z) = z$ for every pure-strategy vertex $z$ of $\bar{\mathcal{M}}$.
\end{definition}

The reparametrisation $\tau$ is needed because the homeomorphism must preserve the direction of time whilst leaving its rate free, allowing the two models to traverse matched orbits at different speeds. The stricter notion of \emph{conjugacy}, which fixes $\tau(z, t) = t$ and therefore preserves the period of every closed orbit, does not hold here. Even in games where topological equivalence holds, such as the matching-pennies family of Corollary~\ref{cor:isochronous}, the log-linear centre cycles with one fixed period, whereas the linear centre's period grows without bound near the boundary. Equivalence in the sense of Definition~\ref{def:topological-equivalence} is therefore the strongest relation that can hold between the two models. Theorem~\ref{thm:partition} establishes where it holds.

A two-player, two-action game is defined by the payoff matrices
$$ \begin{aligned}
A = \begin{bmatrix}
R_A & S_A \\
T_A & P_A
\end{bmatrix}
, \qquad
B = \begin{bmatrix}
R_B & S_B \\
T_B & P_B
\end{bmatrix}
\end{aligned} $$
and is a point $(R_i,S_i,T_i,P_i)_{i\in \{A, B\}} \in \mathbb{R}^{4 \times 2}$. 
In any such game the vector field in log-odds coordinates is orthogonally decoupled: each agent's fitness depends on the opponent's strategy alone, so the velocity $\dot{v} = f^A_1(y) - f^A_2(y)$ depends only on the opponent's coordinate $w$, and $\dot{w}$ only on $v$.
In the probability coordinate the same velocity reads $\dot{x}_1 = x_1(1 - x_1)\,\dot{v}$, since $x_1 = \sigma(v)$ and $\sigma' = \sigma(1 - \sigma)$. The factor $x_1(1 - x_1)$ is strictly positive, so the magnitude of $\dot{x}_1$ depends on $x_1$ but its sign is set by $y_1$ alone on the interior.

The decoupling forces the interior nullclines (where $\dot{v} = 0$, and thus $\dot{x}_1 = 0$) to be horizontal and vertical lines, slicing the state space into rectangular quadrants within which neither velocity component can change sign. The flow direction is therefore fixed within each quadrant, and the one-dimensional flow orientations along the simplex boundaries determine the existence, sign structure, and index (the net number of turns the flow direction makes along a small loop around it) of any interior equilibrium. The index separates saddles (index $-1$) from equilibria around which the flow rotates (index $+1$), but it does not show whether the latter are centres, with closed orbits, or spirals. That requires the Hamiltonian of \S\ref{sec:hamiltonian-structure}.

Under the linear replicator dynamics, the directional flow of a strategy is driven by its comparative expected payoff: $\dot{x}_2 \propto (Ay)_2 - (Ay)_1$. Evaluating this velocity at the boundaries of the opponent's simplex ($y_2 \to 0$ and $y_2 \to 1$) reduces the flow orientation to the algebraic difference between raw payoff parameters. 

\begin{definition}[Boundary incentive differentials]
The raw payoff parameters $(R_i, S_i, T_i, P_i)$ of Agent $i$ induce two boundary incentive differentials $\Delta_{y_2}^i$:
\begin{equation}
    \Delta_0^i = T_i - R_i
    \quad \text{and} \quad
    \Delta_1^i = P_i - S_i
\end{equation}
quantifying $i$'s incentive to play Strategy 2 when the opponent plays pure Strategy 1 ($y_2 \to 0$) and pure Strategy 2 ($y_2 \to 1$), respectively.
\end{definition}

These boundary incentives correspond to the classical game-theoretic motivations of greed and fear \cite{Coombs1973ReparameterizationPrisoners}. Assuming Strategy 1 represents cooperation and Strategy 2 represents defection: $\Delta_0^i = T_i - R_i$ quantifies greed (the incentive to defect against a cooperating opponent), whilst $\Delta_1^i = P_i - S_i$ quantifies fear (the incentive to defect against a defecting opponent). They are also, up to a sign, the two payoffs left by the normalisation that \citeauthor{Weibull1997EvolutionaryGame}~\cite[\S1.5.3]{Weibull1997EvolutionaryGame} uses for symmetric $2 \times 2$ games. Subtracting $T_i$ from the first column of agent $i$'s payoff matrix and $S_i$ from the second is a gauge transformation (\S\ref{sec:gauge}). It sets the off-diagonal entries to zero and leaves the diagonal matrix $\operatorname{diag}(-\Delta_0^i, \Delta_1^i)$.

In the log-odds coordinate $v$, the linear and log-linear velocities are
\begin{align*}
    \dot{v}_\mathrm{Lin} &= -\Delta_0^i y_1 - \Delta_1^i y_2 \\
    \dot{v}_\mathrm{Log} &= -\Delta_0^i \ln y_1 - \Delta_1^i \ln y_2
\end{align*}
The topological comparison rests on the limits of these velocities as the opponent's strategy approaches the boundary.

\begin{theorem}[Partition of bimatrix games into topologically equivalent and topologically non-equivalent classes]\label{thm:partition}
The linear and log-linear flows of a $2 \times 2$ bimatrix game are topologically non-equivalent in the vertex-preserving sense of Definition~\ref{def:topological-equivalence} if and only if at least one agent satisfies the one-dimensional non-equivalence condition
\begin{equation}\label{eqn:preservation-condition}
    \mathcal{P}_i : (\Delta_0^i \cdot \Delta_1^i > 0) \lor (\Delta_0^i \cdot \Delta_1^i = 0 \land \Delta_0^i \ne \Delta_1^i).
\end{equation}
The entire parameter space $\mathbb{R}^{4 \times 2}$ is thereby partitioned into the set $\mathcal{D}$ of topologically non-equivalent games,
\begin{equation*}
    \mathcal{D} = \big\{ (R_i,S_i,T_i,P_i)_{i\in \{A, B\}} \in \mathbb{R}^{4 \times 2} \,\big|\,
    \exists i \in \{A,B\}, \; \mathcal{P}_i  \big\},
\end{equation*}
and its complement, $\mathcal{E} = \mathbb{R}^{4 \times 2} \setminus \mathcal{D}$, containing topologically equivalent flows.

Condition $\mathcal{P}_i$ has a direct game-theoretic meaning. The branch $\Delta_0^i \Delta_1^i > 0$ says that agent $i$ has a strictly dominant pure strategy (Strategy 2 when both differentials are positive, Strategy 1 when both are negative); the branch $\Delta_0^i \cdot \Delta_1^i = 0 \land \Delta_0^i \ne \Delta_1^i$ says that agent $i$ has a weakly but not strictly dominant pure strategy; the excluded case $\Delta_0^i = \Delta_1^i = 0$ is full payoff indifference. Hence
\begin{equation*}
    \mathcal{D} = \left\{\,\parbox{0.7\linewidth}{\centering games in which some agent has a weakly dominant pure strategy, excluding full payoff indifference}\,\right\}.
\end{equation*}
Equivalently, $\mathcal{D}$ is the set of games with no interior rest point under either model. The log-linear substitution has no topological effect wherever an interior equilibrium exists to pin the flow.
\end{theorem}

\begin{proof}[Proof sketch]
The full proof is in Appendix~\ref{apx:partition-proof}. For \emph{necessity}, suppose some agent satisfies $\mathcal{P}_i$; relabel so that it is Agent A and the log-linear flow increases $v$. On every branch of $\mathcal{P}_A$ the two scalar velocities then have constant, opposite signs throughout the interior, so the forward limit sets of the two flows are confined to disjoint parts of the boundary: log-linear orbits end on the face $\{x_1 = 1\}$ or stall towards $\{y_1 = 0\}$, linear orbits on $\{x_1 = 0\}$ or $\{y_1 = 1\}$. A vertex-preserving equivalence transports limit sets onto limit sets (Lemma~\ref{lem:limit-transport}), so a family of log-linear orbits whose limit points approach a vertex forces the continuous extension $\bar{h}$ to map that vertex into a closed boundary set that excludes it. A case analysis over the opponent's parameters shows that such a family always exists, contradicting vertex preservation.

For \emph{sufficiency}, both agents fail $\mathcal{P}_i$, and each flow conserves an energy $H = g_A(w) - g_B(v)$ whose components are antiderivatives of the scalar velocities: strictly convex or strictly concave, coercive, and of the same convexity pattern in both models. Both flows therefore share their global orbit structure, a centre when the agents' coordination types are mixed and a saddle when they match. Transporting each nullcline quadrant to the energy coordinates $\big(|g_B - g_B^*|,\, |g_A - g_A^*|\big)$ and matching level sets across the models gives a homeomorphism that carries orbits to orbits, preserves time orientation, and extends to the closed square fixing every vertex. Fully indifferent agents are handled separately by a skew-product construction, and the characterisations of $\mathcal{D}$ in the statement follow from the sign definitions of weak dominance (Part III of the appendix proof).
\end{proof}

One agent satisfying $\mathcal{P}_i$ is enough to place the game in $\mathcal{D}$. An equivalence must preserve both agents' flow orientations at once, so a coordination motif in Agent A's payoffs cannot prevent the change if Agent B's payoffs satisfy $\mathcal{P}_B$.

\begin{remark}[Vertex preservation carries the content on strict dominance]\label{rem:parallelisable}
On the strict-dominance interior of $\mathcal{D}$ ($\Delta_0^A \Delta_1^A > 0$), both scalar velocities are bounded away from zero, so every orbit of either flow crosses the line $\{v = 0\}$ exactly once. Each flow admits a global cross-section (a curve that every orbit crosses exactly once), is parallelisable (its orbits can be straightened into parallel lines), and is topologically equivalent to the constant flow on $\mathbb{R}^2$, hence to the other, by an \emph{unrestricted} homeomorphism. Without the vertex-preserving requirement of Definition~\ref{def:topological-equivalence}, the non-equivalence set $\mathcal{D}$ would lose its strict-dominance interior entirely; equally, this is why the necessity proof in Appendix~\ref{apx:partition-proof} must argue through boundary limit sets (the boundary points that orbits approach as time runs forwards or backwards) rather than any interior invariant, no weaker instrument being able to distinguish the flows. (On the weak-dominance boundary of $\mathcal{D}$, orbits may stall and the cross-section argument does not apply.)
\end{remark}

\begin{corollary}[Taxonomy of games]\label{cor:taxonomy}
Standard $2 \times 2$ game classes can be categorised as follows.

\textbf{Set $\mathcal{D}$ (topologically non-equivalent flows)} contains every strict-dominance game (Prisoner's Dilemma, Deadlock, and Harmony among them): For any single agent $i$ with a dominant pure strategy, the boundary incentives point in the same direction: $\Delta^i_0$ and $\Delta^i_1$ share the same sign. Therefore, their product is positive: $\Delta^i_0 \cdot \Delta^i_1 > 0$. Any game containing a dominant strategy belongs to $\mathcal{D}$; conversely, by the characterisation in Theorem~\ref{thm:partition}, every game in $\mathcal{D}$ contains a weakly dominant pure strategy for some agent.

\textbf{Set $\mathcal{E}$ (topologically equivalent flows)} consists of
    (1) coordination games (Stag Hunt): For both agents,
    $R > T \implies \Delta_0 < 0$ and
    $P > S \implies \Delta_1 > 0$, making $\Delta_0 \cdot \Delta_1 < 0$. Equivalence holds. The interior saddle is preserved.
    (2) anti-coordination games (Chicken): For both agents,
    $T > R \implies \Delta_0 > 0$ and 
    $S > P \implies \Delta_1 < 0$, making $\Delta_0 \cdot \Delta_1 < 0$. Equivalence holds. The interior saddle is preserved.
    (3) cyclic games (Matching Pennies): Agent A seeks to coordinate:
    $R_A > T_A \implies \Delta_0^A < 0$ and 
    $P_A > S_A \implies \Delta_1^A > 0$, so 
    $\Delta_0^A \cdot \Delta_1^A < 0$. Agent B seeks to anti-coordinate: 
    $T_B > R_B \implies \Delta_0^B > 0$ and 
    $S_B > P_B \implies \Delta_1^B < 0$, so 
    $\Delta_0^B \cdot \Delta_1^B < 0$. Equivalence holds. The conditions for preservation of the interior centre are analysed in \S\ref{sec:hamiltonian-structure}.
\end{corollary}

The partition also has a static characterisation, with no dynamics in its statement: Condition $\mathcal{P}_i$ is the obstruction to representing agent $i$'s \emph{linear} fitness as conventional Bayesian evidence in the sense of \S\ref{sec:bayesian-evidence}.

\begin{proposition}[Conventional representability is the complement of $\mathcal{P}_i$]\label{prop:representability}
For a two-action agent with payoffs $(R, S, T, P)$, there exists $c \in \mathbb{R}^2$ such that both rows of $A + \mathbf{1}c^{\top}$ are entrywise logarithms of probability vectors, rendering the linear fitness, modulo the gauge, an expected log-likelihood over opponent actions, if and only if the agent fails Condition $\mathcal{P}_i$: either $\Delta_0 \Delta_1 < 0$, or $\Delta_0 = \Delta_1 = 0$. Consequently $\mathcal{D}$ is exactly the set of games in which some agent's linear fitness admits no conventional Bayesian representation, and on $\mathcal{E}$ every agent's does.
\end{proposition}

\begin{proof}
The representation requires $u = (e^{c_1}, e^{c_2})$ with $M u = \mathbf{1}$, $u > 0$, where $M = \exp(A)$ entrywise. If $\Delta_0 = \Delta_1 = 0$ the two rows of $M$ coincide and the single constraint $e^{R} u_1 + e^{S} u_2 = 1$ admits strictly positive solutions. If $\Delta_0 = \Delta_1 \ne 0$ the second row of $M$ is $e^{\Delta_0}$ times the first, so $Mu = \mathbf{1}$ would force $1 = e^{\Delta_0}$: the system is inconsistent and no solution exists, positive or otherwise. If $\Delta_0 \ne \Delta_1$, then $\det M = e^{R+P} - e^{S+T}$ is non-zero and carries the sign of $\Delta_1 - \Delta_0$, and Cramer's rule gives $u_1 = (e^{P} - e^{S})/\det M$ and $u_2 = (e^{R} - e^{T})/\det M$, with numerator signs $\operatorname{sign}(\Delta_1)$ and $\operatorname{sign}(-\Delta_0)$. If $\Delta_0 \Delta_1 < 0$, the three signs $\operatorname{sign}(\Delta_1)$, $\operatorname{sign}(-\Delta_0)$, $\operatorname{sign}(\Delta_1 - \Delta_0)$ coincide and $u > 0$. If $\Delta_0 \Delta_1 > 0$, the two numerators have opposite signs, so no positive solution exists for either sign of the determinant. If exactly one differential vanishes, $M$ is invertible and the unique solution has a zero component, a point mass excluded by the logarithm.
\end{proof}

Together with Proposition~\ref{prop:conventional-obstruction}, which supplies the general $n \times m$ obstruction, this proposition establishes that the games on which the linear model loses its conventional Bayesian credentials are, for two actions, exactly the games on which the two flows are topologically non-equivalent (Theorem~\ref{thm:partition}), closing the loop opened in \S\ref{sec:bayesian-evidence}. The mismatch is localised on $\mathcal{D}$, where the log-linear representation inverts dominance.

Proposition~\ref{prop:representability} is the case $m = 2$ of Corollary~\ref{cor:representability-2xm}, since the difference of the two payoff rows is $(R - T, S - P) = (-\Delta_0, -\Delta_1)$. The coincidence between representability and the existence of an interior rest point does not extend beyond two actions. In the $3 \times 3$ game of Remark~\ref{rem:no-dominance-insufficient}, the uniform opponent strategy gives all three of the agent's pure strategies the same expected payoff, which is the agent's condition for an interior rest point of the linear flow, yet no conventional representation exists.

\begin{figure}[tbp]
    \centering
    \includegraphics[width=\linewidth]{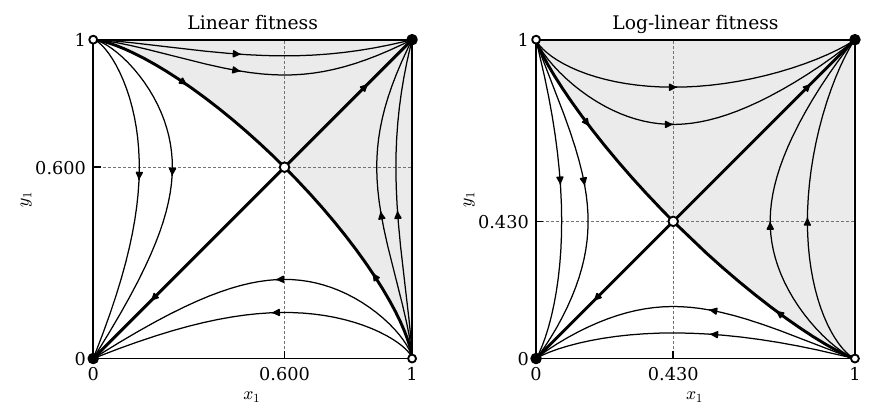}
    \caption{The Stag Hunt of Remark~\ref{rem:displacement}, $(R, S, T, P) = (5, 0, 3, 3)$ for both agents. Strategy 1 is Stag. The game lies in Set $\mathcal{E}$. The interior saddle and its separatrix (thick, running between the off-diagonal vertices) are preserved, and the invariant diagonal is the unstable manifold (the orbits that leave the saddle). The shaded region is the basin of the payoff-dominant equilibrium at $(1,1)$. The substitution displaces the saddle from $y_1 = 0.600$, the mixed Nash equilibrium, to $y_1 \approx 0.430$: the basin boundary crosses $1/2$, so the payoff-dominant action attracts the majority of initial conditions whilst Hare remains risk-dominant.}
    \label{fig:sh-displacement}
\end{figure}

\begin{remark}[Equilibrium displacement on $\mathcal{E}$]\label{rem:displacement}
While topological equivalence preserves the orbit structure, it does not preserve the location of rest points. On $\mathcal{E}$, for an agent with $\Delta_0^i \Delta_1^i < 0$ (a fully indifferent agent has no isolated nullcline, and this remark does not concern it), the interior nullcline sits at
\begin{align*}
    y_1^{\mathrm{Lin}} &= \frac{\Delta_1^i}{\Delta_1^i - \Delta_0^i} && \text{(linear)}, \\
    y_1^{-\Delta_0^i}\, y_2^{-\Delta_1^i} &= 1 && \text{(log-linear)},
\end{align*}
and the two generically (for almost all payoffs) differ. Write $r_i = \Delta_1^i / (-\Delta_0^i)$, which is strictly positive on this stratum. The linear nullcline is then $y_1^{\mathrm{Lin}} = r_i / (1 + r_i)$, and substituting it into the log-linear nullcline equation, in the equivalent form $y_1^{-\Delta_0^i} = y_2^{\Delta_1^i}$, reduces after taking logarithms to the scalar condition $r_i = (1 + r_i)^{1 - r_i}$. The function $g(r) = \ln r - (1-r)\ln(1+r)$ has derivative $g'(r) = 1/r + \ln(1+r) - (1-r)/(1+r) > 0$ for every $r > 0$, so $g$ is strictly increasing and its only zero is $r_i = 1$. The two models therefore place agent $i$'s interior nullcline at the same location if and only if
$$\Delta_0^i = -\Delta_1^i,$$
in which case the shared rest point is the barycentre $y_1^* = 1/2$. The coincidence set is thus a measure-zero hyperplane in payoff space, and is the zero locus of the \emph{intrinsic bias} $\eta_i = -\tfrac{1}{4}(\Delta_0^i + \Delta_1^i)$ (\S\ref{sec:interaction-curvature}), the same regime in which the dynamics \eqref{eq:scalar-logit-system} linearise and the zero-sum cycles become isochronous (Corollary~\ref{cor:isochronous}). Displacement of the interior equilibrium is the signature of this bias.

In the symmetric Stag Hunt with $(R, S, T, P) = (5, 0, 3, 3)$, so that $\Delta_0 = -2$, $\Delta_1 = 3$ and hence $\eta = -1/4 \neq 0$, the linear rest point is the mixed Nash equilibrium at $y_1^{\mathrm{Nash}} = 0.600$, whereas the log-linear rest point solves $y_1^{2} = (1 - y_1)^{3}$, giving $y_1^{\dagger} \approx 0.430$. Under log-linear dynamics, $43\%$ confidence that the opponent hunts stag suffices to commit to stag, rather than $60\%$ (Figure~\ref{fig:sh-displacement}).

The displacement here crosses $1/2$, which changes the selection ranking. Risk dominance in the sense of \citeauthor{Harsanyi1988GeneralTheory}~\cite{Harsanyi1988GeneralTheory} is a property of the payoff matrix, not of the dynamics, and is therefore left untouched by the substitution. Hare remains risk-dominant, since $|\Delta_1| = 3 > |\Delta_0| = 2$. Under the linear model the static and dynamic criteria agree, because $y_1^{\mathrm{Nash}} = 0.600 > 1/2$ assigns the larger basin to the risk-dominant action, in accordance with the stochastic-stability selection results of \citeauthor{Kandori1993LearningMutation}~\cite{Kandori1993LearningMutation} and \citeauthor{Young1993EvolutionConventions}~\cite{Young1993EvolutionConventions}. Under the log-linear model the basin boundary sits at $y_1^{\dagger} \approx 0.430 < 1/2$, so the \emph{payoff}-dominant action attracts the majority of initial conditions while the risk-dominance ranking is unchanged: on this side of the crossing, dynamic equilibrium selection and risk dominance disagree.

Both basin-majority claims follow from the position of the separatrix (the curve that divides the two basins of attraction) relative to the anti-diagonal $x_1 + y_1 = 1$, where $(x_1, y_1)$ are the two agents' probabilities of Stag. Under the log-linear model $\dot{x}_1 = x_1(1 - x_1)\,\varphi(y_1)$ with $\varphi(u) = 2 \ln u - 3 \ln(1 - u)$, and symmetrically for $\dot{y}_1$, so on the anti-diagonal we have 
$$\frac{d}{dt}(x_1 + y_1) = x_1 y_1 \big[\varphi(x_1) + \varphi(y_1)\big] = -\,x_1 y_1 \ln(x_1 y_1) > 0$$
and every trajectory crosses the anti-diagonal upward. The saddle lies strictly below that line, and a trajectory that meets the line is carried above it for good, so the stable manifold of the saddle (the states whose orbits end at the saddle) never meets its open segment. The separatrix therefore lies below the anti-diagonal, touching it only at the off-diagonal vertices, and the basin of mutual Stag contains the open triangle above the anti-diagonal together with a band below it, i.e. more than half of the square. The same computation under the linear model gives $\varphi_{\mathrm{Lin}}(u) = 5u - 3$, hence $\frac{d}{dt}(x_1 + y_1) = -\,x_1 y_1 < 0$ on the anti-diagonal, with the saddle strictly above it. In this case the containment reverses, and the risk-dominant equilibrium attracts the majority.

Two consequences follow. First, the Folk Theorem recalled at the opening of this section fails for the log-linear model: its interior rest points are generically not Nash equilibria of the underlying game. Second, on $\mathcal{E}$ the substitution preserves the topology but moves the mixed equilibrium, the basin boundary, and the induced risk-dominance threshold; here they move far enough that dynamic selection and risk dominance disagree.

One caveat applies to these interior quantities. They are computed for the fitness \eqref{eqn:log-linear-fitness} as written, which scores an ordered sequence of opponent actions rather than their multinomial counts (\S\ref{sec:bayesian-evidence}). Restoring the omitted combinatorial coefficient would add a constant to each payoff row, and the log-linear nullclines would move, taking $y_1^{\dagger}$ and the crossing with them. The partition of Theorem~\ref{thm:partition}, on the other hand, is driven by the boundary divergences, which no finite row constant alters. In games with a dominant strategy, however, a large enough constant can also create new interior nullclines and rest points (Remark~\ref{rem:multinomial-coefficient}).
\end{remark}

The topological non-equivalence of strict dominance games under the log-linear model is driven by unbounded log-loss. As the opponent's mixture approaches certainty, the rate of log-evidence accumulation diverges. A strategy row that places weight on the vanishing action is a hypothesis whose log-likelihood diverges. The fitness difference between the two rows diverges at a rate set by their weight difference on that action, with sign fixed by the boundary incentive differentials. This is the likelihood-side counterpart of Cromwell's Rule in Bayesian inference, which bars prior probabilities of exactly zero (or one) because they preclude any future updating of those probabilities \cite{Lindley1985MakingDecisions}. Here the divergence appears in the likelihood rather than the prior, accruing unbounded negative evidence against any hypothesis that assigns vanishing probability to what is observed. The divergence does not preserve interior beliefs, as trajectories still converge to boundary states and eliminate a hypothesis. Instead, near certainty, the divergence redirects \textit{which} hypothesis is eliminated, inverting the flow rather than arresting it. Bounded precision would introduce finite limits through entropy regularisation, but the boundary structure of the log-linear model remains distinct from the linear model at every interior state. The inversion is not an artifact of the boundary limit.

In a strict-dominance game, the two logarithmically diverging terms at each edge of the opponent's simplex act against the dominant strategy. The relative velocity of cooperation over defection evaluates to $f_C - f_D = -\Delta_0 \ln y_C - \Delta_1 \ln y_D$. Because both opponent probabilities $y \in (0,1)$ in the interior, their natural logarithms are negative. In a Prisoner's Dilemma ($\Delta_0 > 0, \Delta_1 > 0$), the velocity is a sum of two positive terms, so $f_C - f_D$ is strictly positive everywhere in the interior of the simplex.

\begin{figure}[tbp]
    \centering
    \includegraphics[width=\linewidth]{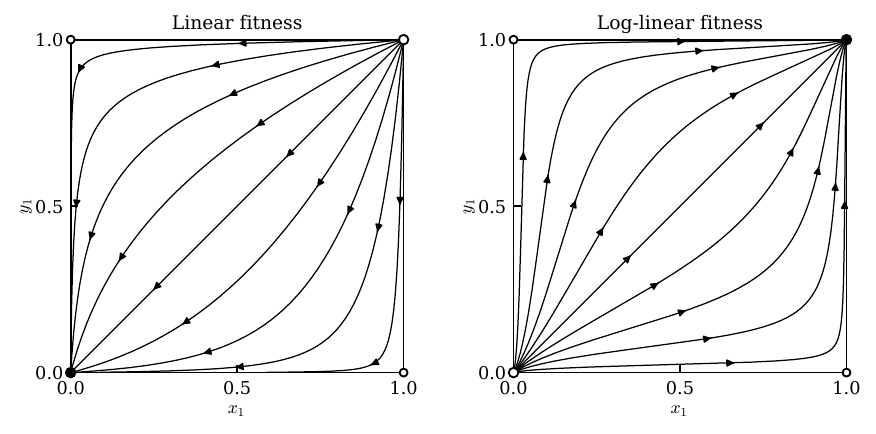}
    \caption{Exact phase portraits of the Prisoner's Dilemma with $(R, S, T, P) = (3, 0, 5, 1)$ for both agents; Strategy 1 is Cooperate, and orbits are level sets of the conserved quantity of each model (Appendix~\ref{apx:partition-proof}, Step 4). Under linear fitness every interior trajectory converges to mutual defection at $(0,0)$. Under log-linear fitness the dominant strategy is inverted (Proposition~\ref{prop:dominance-inversion}) and every interior trajectory converges to mutual cooperation at $(1,1)$. Filled circles mark attractors; open circles mark repelling or saddle rest points.}
    \label{fig:pd-inversion}
\end{figure}

As the opponent's strategy approaches certainty, the log-evidence diverges. The incentives that make defection strictly dominant in the linear model, $T > R$ and $P > S$, also make the divergence against defection dominate the divergence against cooperation. The dominant strategy is inverted in every strict-dominance game---this is the two-action instance of the general Proposition~\ref{prop:dominance-inversion}. Corollary~\ref{cor:taxonomy} places the Harmony game in Set $\mathcal{D}$, where $\Delta_0 < 0$ and $\Delta_1 < 0$. Under the exact scalar velocity $\dot{v}_\mathrm{Log} = -\Delta_0 \ln y_1 - \Delta_1 \ln y_2$, both logarithmic terms are negative, so $\dot{v}_\mathrm{Log} < 0$ everywhere. Hence every interior orbit of the Harmony game, a game of common interest in which mutual cooperation is dominant, converges to mutual defection: the log-linear flow carries every interior orbit of the Prisoner's Dilemma to mutual cooperation and every interior orbit of the Harmony game to mutual defection. Figure~\ref{fig:pd-inversion} draws the two phase portraits.

\subsection{Interaction Curvature and Bias}\label{sec:interaction-curvature}

The scalar velocities take a two-parameter form in the coordinates $(v, w)$. Write $s = \ln y_1 + \ln y_2$ for the opponent's summed log-probabilities, so that $\ln y_1 = (s + w)/2$ and $\ln y_2 = (s - w)/2$. Substituting into the log-linear velocity of \S\ref{sec:topological-equivalence} gives
$$
\dot{v} = -\Delta_0^A \ln y_1 - \Delta_1^A \ln y_2
        = \tfrac{1}{2}\left(\Delta_1^A - \Delta_0^A\right) w - \tfrac{1}{2}\left(\Delta_0^A + \Delta_1^A\right) s .
$$
Because $y_1 = \sigma(w)$, the sum $s$ is a function of $w$ alone,
$$
\tilde{s}(w) = \ln\big(y_1 y_2\big) = w - 2\ln(1 + e^{w}),
$$
and the coupled system becomes
\begin{equation}\label{eq:scalar-logit-system}
\begin{aligned}
    \dot{v} &= 2\mu_A w + 2\eta_A \tilde{s}(w) \\
    \dot{w} &= 2\mu_B v + 2\eta_B \tilde{s}(v)
\end{aligned}
\end{equation}
with, for each agent,
\begin{equation}\label{eqn:mu-eta}
    \mu = \tfrac{1}{4}\left(\Delta_1 - \Delta_0\right),
    \qquad
    \eta = -\tfrac{1}{4}\left(\Delta_1 + \Delta_0\right).
\end{equation}
These are the two parameters of the dynamics. We call $\mu$ the \emph{interaction curvature}: it measures how the agent's incentive changes with the opponent's mixture. We call $\eta$ the \emph{intrinsic bias}: a constant pull towards the first strategy ($\eta < 0$) or the second ($\eta > 0$). The bias pulls towards cooperation in the Prisoner's Dilemma, where $T > R > P > S$.

When both agents' biases vanish ($\eta_A = \eta_B = 0$), the system \eqref{eq:scalar-logit-system} is linear and has explicit solutions. This is the class studied by \citeauthor{Ay2005NotionLinear}~\cite{Ay2005NotionLinear}, for a single population. If either bias is non-zero, its term $\tilde{s}$ makes the system non-linear and that theory does not apply.

\subsection{Hamiltonian Structure}\label{sec:hamiltonian-structure}

Hofbauer~\cite{Hofbauer1996EvolutionaryDynamics} proved that the \emph{linear} replicator dynamics of bimatrix games are Hamiltonian. More recently, \citeauthor{Ota2025HamiltonianPolymatrix}~\cite{Ota2025HamiltonianPolymatrix} gave a Hamiltonian for follow-the-regularised-leader learning (each agent plays a regularised best response to its cumulative payoffs), which includes the linear replicator dynamics, in networks of pairwise zero-sum games. Their canonical variables are the strategies and the cumulative payoffs, and the construction relies on the zero-sum property. Here we prove the analogue of Hofbauer's result for the log-linear dynamics and connect it to the topological partition above. The canonical coordinates are the agents' log-odds, and the result holds for every two-action game, zero-sum or not.

The log-linear dynamics of any $2 \times 2$ bimatrix game are the coupled system \eqref{eq:scalar-logit-system}. Each velocity depends only on the opponent's coordinate, so the Jacobian has zero trace for every game in the class:
$$\nabla \cdot \mathcal{X}_{\mathrm{Log}} = \frac{\partial \dot{v}}{\partial v} + \frac{\partial \dot{w}}{\partial w} = 0 + 0 = 0.$$
The flow therefore preserves area in the $(v, w)$ plane, and no interior rest point can be asymptotically stable. A divergence-free field on the plane admits a global stream function (a function whose level curves are the orbits), so a Hamiltonian exists. The proposition below shows its closed form, which contributes to the Hessian classification, identifies the zero-sum invariant, and shows that no choice of payoffs produces an interior attractor.

\begin{proposition}[Closed-form Hamiltonian for the log-linear dynamics]\label{prop:hamiltonian}
The log-linear dynamics \eqref{eq:scalar-logit-system} of any $2 \times 2$ bimatrix game admit the global conserved quantity
$$H(v, w) = \mu_A w^2 - \mu_B v^2 + 2\eta_A \tilde{S}(w) - 2\eta_B \tilde{S}(v),$$
which is a Hamiltonian for the flow in the canonical coordinates $(v, w)$ (that is, $\dot{v} = \partial H/\partial w$ and $\dot{w} = -\partial H/\partial v$), where $\tilde{S}(z) = \int \tilde{s}(z)\, dz$ denotes an antiderivative of the sum-of-log-probabilities function $\tilde{s}$; explicitly,
$$\tilde{S}(z) = \frac{z^2}{2} + 2\operatorname{Li}_2(-e^{z}),$$
where $\operatorname{Li}_2$ is the dilogarithm (a standard special function), since $\tfrac{d}{dz}\operatorname{Li}_2(-e^{z}) = -\ln(1 + e^{z})$. The additive constant of integration leaves the level sets of $H$ unchanged.
\end{proposition}

\begin{proof}
Direct computation gives $\partial H/\partial w = 2\mu_A w + 2\eta_A \tilde{s}(w) = \dot{v}$ and $-\partial H/\partial v = 2\mu_B v + 2\eta_B \tilde{s}(v) = \dot{w}$, confirming the canonical equations $\dot{v} = \partial H/\partial w$ and $\dot{w} = -\partial H/\partial v$.
\end{proof}

In particular, the first integral (conserved quantity) constructed in Step 4 of the sufficiency proof (Appendix~\ref{apx:partition-proof}) is recovered here in closed form: the decomposition $H(v, w) = g_A(w) - g_B(v)$ holds with
$$g_A(w) = \mu_A w^2 + 2\eta_A \tilde{S}(w), \qquad g_B(v) = \mu_B v^2 + 2\eta_B \tilde{S}(v).$$
Furthermore, since $\tilde{s}'(w) = y_2 - y_1$, the sign-constant quantity of that proof satisfies $g_A''(w) = 2\mu_A + 2\eta_A\, \tilde{s}'(w) = \Delta_1^A y_1 - \Delta_0^A y_2$, being twice the effective interaction curvature $\hat{\mu}_A$ introduced below, evaluated along the whole line rather than at the equilibrium.

The Hessian of $H$ at an interior fixed point $(v^*, w^*)$ is the diagonal matrix $\operatorname{diag}(-2\hat{\mu}_B,\; 2\hat{\mu}_A)$, where $\hat{\mu}_A = \mu_A + \eta_A \tilde{s}'(w^*)$ and $\hat{\mu}_B = \mu_B + \eta_B \tilde{s}'(v^*)$ are the \emph{effective interaction curvatures} evaluated at the equilibrium. Because $\tilde{s}'(w) = y_2 - y_1$, substituting \eqref{eqn:mu-eta} gives $\hat{\mu}_A = \frac{1}{2}(\Delta_1^A y_1^* - \Delta_0^A y_2^*)$. On the complement set $\mathcal{E}$ (where $\Delta_0^A \Delta_1^A < 0$), for a coordination-type agent ($\Delta_0^A < 0 < \Delta_1^A$), both terms in $\hat{\mu}_A$ are positive. For an anti-coordination agent ($\Delta_0^A > 0 > \Delta_1^A$), both terms are negative. Consequently, $\hat{\mu}_A$ shares the sign of $\Delta_1^A - \Delta_0^A$ on $\mathcal{E}$. The bias $\eta$ alters the magnitude of the effective curvature but cannot invert its sign. Thus, matching coordination types strictly yield a saddle ($\hat{\mu}_A \hat{\mu}_B > 0$) and mixed types strictly yield a centre ($\hat{\mu}_A \hat{\mu}_B < 0$). The Hessian's signature defines a three-way geometric partition:
\begin{itemize}
    \item $\hat{\mu}_A \hat{\mu}_B > 0$ (every Stag Hunt, Chicken): The Hessian is indefinite. Level sets near $(v^*, w^*)$ are hyperbolas; trajectories are open, and the fixed point is a saddle.
    \item $\hat{\mu}_A \hat{\mu}_B < 0$ (every Matching Pennies): The Hessian is definite. Level sets are ellipses centred at the fixed point, generating closed periodic orbits. The fixed point is a conservative centre.
    \item $\hat{\mu}_A \hat{\mu}_B = 0$: The Hessian is singular. Recall that $\hat{\mu}_A = \frac{1}{2}(\Delta_1^A y_1^* - \Delta_0^A y_2^*)$. For this to equal zero at an interior equilibrium (where the velocity $-\Delta_0^A \ln y_1^* - \Delta_1^A \ln y_2^* = 0$), it requires the condition $y_1^* \ln y_1^* + y_2^* \ln y_2^* = 0$. Because the negative entropy is strictly less than zero everywhere on the open simplex, the condition cannot hold. A singular Hessian therefore arises only in the fully degenerate case $\Delta_0 = \Delta_1 = 0$ for at least one agent, generating a continuous manifold of equilibria rather than an isolated fixed point. Games with non-zero equal incentives ($\Delta_0 = \Delta_1 \neq 0$) satisfy condition $\mathcal{P}_i$ \eqref{eqn:preservation-condition} and have no interior fixed point, so the Hessian analysis does not apply to them.
\end{itemize}

For unbiased zero-sum games ($B = -A^{\top}$, $\eta = 0$), the identity $\mu_B = -\mu_A$ reduces the Hamiltonian to a squared Euclidean distance in the $(v, w)$ plane, $H(v, w) \propto \|v - v^*\|^2 + \|w - w^*\|^2$. The orbits are circles centred on the equilibrium. They correspond to closed periodic orbits on the simplex because the log-odds map is a homeomorphism, preserving the interior centre on Set $\mathcal{E}$. The invariant differs from that of the \emph{linear} zero-sum replicator dynamics, whose conserved quantity is the sum of Kullback--Leibler divergences to the equilibrium, $\KL{x^*}{x} + \KL{y^*}{y}$, due to \citeauthor{Akin1984EvolutionaryDynamics}~\cite{Akin1984EvolutionaryDynamics} (see \citeauthor{Sato2003CoupledReplicator}~\cite{Sato2003CoupledReplicator} for its role in learning dynamics). The dynamics under the two models have different level sets, and the Kullback--Leibler sum is not conserved by the log-linear flow. In log-odds coordinates the sum equals $\ln \cosh(v/2) + \ln \cosh(w/2)$, which is not constant along the circular orbits.

Such conservative centres are structurally unstable (due to the equations rather than of the trajectories). The orbits close because the flow neither damps nor drives the cycling, the eigenvalues at the fixed point being purely imaginary, and an arbitrarily small perturbation of the vector field can therefore tip every closed orbit into a spiral. Payoff perturbations cannot do this, because the zero trace holds for every payoff matrix in the class. Perturbations that add self-dependence to the velocities, such as an entropy-regularisation term or informational noise, do break the conservation, turning the centre into an attractor or a repeller.

\begin{corollary}[Isochronous zero-sum cycling]\label{cor:isochronous}
For an unbiased zero-sum game ($\eta_A = \eta_B = 0$ and $B = -A^{\top}$, so that $\mu_B = -\mu_A \ne 0$; equivalently $T_A = S_A$ and $P_A = R_A$, a two-parameter family containing Matching Pennies), the system \eqref{eq:scalar-logit-system} reduces to an exact harmonic oscillator,
$$\dot{v} = 2\mu_A w, \qquad \dot{w} = -2\mu_A v, \qquad \text{hence} \quad \ddot{v} = -4\mu_A^2\, v,$$
with period
$$T = \frac{\pi}{|\mu_A|}$$
independent of amplitude, i.e., the log-linear centre is \emph{isochronous}. The Hamiltonian of the linear replicator $H_\mathrm{Lin}$ (see the Remark below) is built from softplus rather than quadratic terms, so its period varies with amplitude and diverges as orbits approach the boundary (i.e., the system is not isochronous). This provides a way to discriminate between the two models behaviourally: a log-evidence integrator in Matching Pennies cycles at the fixed rate $\pi / |\mu_A|$ no matter how extreme its initial beliefs, whereas an expected-utility integrator slows down near the boundary.
\end{corollary}

\begin{figure}[tbp]
    \centering
    \includegraphics[width=\linewidth]{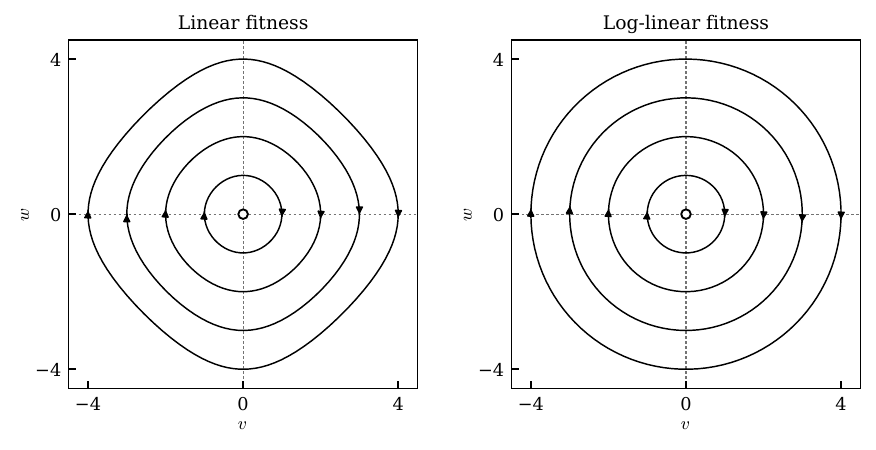}
    \caption{Matching Pennies ($B = -A^{\top}$, unbiased) in log-odds coordinates, drawn for $R_A = P_A = 1$, $S_A = T_A = -1$, so $\mu_A = 1 > 0$ and both flows circulate clockwise about the barycentre (for $\mu_A < 0$ both reverse). Orbits are level sets of the Hamiltonian of each model. The invariant curves of the linear flow are level sets of the softplus-built $H_{\mathrm{Lin}}$ (Remark~\ref{rem:linear-logit}), with period growing without bound towards the boundary. The invariant curves of the log-linear flow are Euclidean circles, each traversed in the same period $\pi / |\mu_A|$ (Corollary~\ref{cor:isochronous}).}
    \label{fig:mp-dual}
\end{figure}

\noindent
Figure~\ref{fig:mp-dual} draws the two invariant-curve families in log-odds coordinates.

\begin{remark}[Contrast with the linear replicator in logit coordinates]\label{rem:linear-logit}
    The Hamiltonian above must not be confused with the Hamiltonian of the linear replicator expressed in logit coordinates, which involves the softplus function. To see the distinction, transform the standard linear dynamics $\dot{x}_1 = x_1(1-x_1)[(Ay)_1 - (Ay)_2]$ to the logit variable $v = \ln(x_1 / x_2)$. Because $\dot{v} = (Ay)_1 - (Ay)_2$ and $y_1 = \sigma(w)$, one obtains
    $$\dot{v} = -\Delta_1^A + 4\mu_A\,\sigma(w), \qquad \sigma(w) = \frac{1}{1+e^{-w}},$$
    whose Hamiltonian is $H_{\mathrm{Lin}}(v, w) = 4\mu_A \ln(1 + e^w) - \Delta_1^A w - 4\mu_B \ln(1 + e^v) + \Delta_1^B v$, as verified directly from the canonical equations $\dot{v} = \partial H_{\mathrm{Lin}}/\partial w$ and $\dot{w} = -\partial H_{\mathrm{Lin}}/\partial v$. This velocity is bounded and saturating: as $w \to \pm\infty$, $\dot{v}$ approaches the finite limits $-\Delta_0^A$ (as $w \to +\infty$) and $-\Delta_1^A$ (as $w \to -\infty$) respectively. The log-linear model, by contrast, has $\dot{v} = 2\mu_A w + 2\eta_A \tilde{s}(w)$ which is unbounded. This difference is the mechanism underlying the topological partition of Theorem~\ref{thm:partition}: for dominant-strategy games where $\Delta_0^A$ and $\Delta_1^A$ share the same sign, the log-linear model's unbounded growth acts as an infinite penalty at the edges of the simplex. This forces an inversion of the vector field at both boundaries that does not occur under the linear model and changes the game's global phase portrait.
\end{remark}

More players change the picture, though not the trace. Consider $N > 2$ agents with two actions each. As long as each agent's fitness depends only on the other agents' states, as it does for the linear fitness and for any extension of \eqref{eqn:log-linear-fitness} that scores the opponents' behaviour, an agent's log-odds velocity does not depend on its own log-odds. Every diagonal entry of the Jacobian is then zero, so its trace vanishes for linear and log-linear fitness alike. The flow therefore still preserves volume in log-odds coordinates, and no interior rest point can be asymptotically stable. What is lost is the planar structure. The nullclines no longer slice the state space into boxes on which each velocity keeps its sign, a divergence-free field in more than two dimensions need not admit a Hamiltonian, and the planar arguments behind Theorem~\ref{thm:partition} and Proposition~\ref{prop:hamiltonian} do not carry over. How the payoff tensor enters the log-linear fitness also depends on how \eqref{eqn:log-linear-fitness} is extended to $N$ players, which this paper leaves open.

\section{Conclusion}

Replacing the linear mixture of probabilities $f(y) = A y$ with the log-linear fitness $f(y) = A \ln y$ aligns the replicator's growth-rate update with the additive arithmetic of Bayesian log-evidence. The conventional expected-log-likelihood representation of the linear model, by contrast, exists for a two-action agent if and only if that agent has no weakly dominant strategy (Proposition~\ref{prop:representability}, and Corollary~\ref{cor:representability-2xm} for any number of opponent actions). In games where the substitution changes the global phase portrait, the linear model cannot be framed as Bayesian updating. Beyond two actions, the representation exists if and only if no mixed strategy weakly dominates another in the exponentiated payoffs $\exp(A)$ (Proposition~\ref{prop:representability-nxm}). Pure dominance still obstructs it (Proposition~\ref{prop:conventional-obstruction}), but its absence does not suffice: the representation fails for almost every game in which the agent has more strategies than its opponent, and for some square games with no dominated strategy of any kind (Remark~\ref{rem:no-dominance-insufficient}).
The substitution has three structural consequences. In log-odds coordinates, the dynamics of every two-action game reduce to the two-parameter form \eqref{eq:scalar-logit-system}, linear except for the bias terms. The Jacobian of that system has zero trace for every payoff matrix, so interior rest points can only be centres or saddles. The linear and log-linear flows are topologically distinct, with no vertex-preserving homeomorphism mapping one onto the other, in every bimatrix game whose boundary incentives share a sign.

This includes strict dominance games. In these, the inversion comes from the infinite penalty that logarithmic evidence evaluation places on hypothesis elimination. The penalty does not prevent belief collapse under any payoff matrix in the class (\S\ref{sec:hamiltonian-structure}). Instead, it redirects the collapse away from the originally dominant equilibrium. 

The Prisoner's Dilemma, being in this class of games, may bear on multi-agent learning, subject to the caveat stated below. Under linear replicator dynamics the Prisoner's Dilemma converges to the strictly dominant strategy of mutual defection from every interior state. Under the log-linear model, the unregularised Bayesian dynamics invert the dominance ordering of every agent that has a dominant strategy (Proposition~\ref{prop:dominance-inversion}), eliminating that strategy when the dominance is strict (Corollary~\ref{cor:strict-elimination}). In the Prisoner's Dilemma both boundary incentives ($\Delta_0 > 0$ and $\Delta_1 > 0$) contribute divergences against the defecting action. Mutual defection becomes unstable, and every interior trajectory reaches the boundary attractor at mutual cooperation. The same inversion carries the Harmony game to mutual defection.

One further constraint bears on use of the model as a learning rule: the zero-trace property of \S\ref{sec:hamiltonian-structure} holds for \emph{every} payoff matrix in the log-linear class, so no choice of payoffs produces an interior attractor. Convergence to interior mixed behaviour requires leaving the class, for instance through entropy regularisation or informational noise, never a different choice of game.

Our results suggest a candidate mechanism for the emergence and destruction of cooperation in multi-agent learning: to the extent that a learning rule's continuous-time mean dynamics are log-evidence-based, the inversion that produces cooperation in social dilemmas carries games of common interest to mutual defection. This connection is conditional. Its linear counterpart is well established (cross-learning and Q-learning variants have replicator-type mean dynamics \cite{Borgers1997LearningReinforcement, Tuyls2003SelectionMutation, Bloembergen2015EvolutionaryDynamics}), but the present model involves no reward sampling, exploration, or discounting, and identifying multi-agent reinforcement learning algorithms whose mean dynamics realise the log-linear replicator remains an open question.

The unregularised models presented reach their stable cooperative sink only at the absolute boundary, whereas cooperation between learning agents typically emerges probabilistically in the interior of the strategy simplex, in mixed strategies sustained by noise and bounded rationality. Our models assume infinite precision and integrate evidence from a non-stationary opponent without temporal discounting.

A natural next step is to connect the model with bounded rationality and thermodynamic dissipation. \citeauthor{Sato2003CoupledReplicator}~\cite{Sato2003CoupledReplicator} introduced an explicit memory-loss (entropy-regularisation) term into the replicator equation, making the learning dynamics dissipative. In log-odds coordinates this term is a linear damping that breaks the zero-trace constraint. In our framework it would act as an inverse temperature (precision) parameter, a linear restoring force drawing the stable cooperative attractor from the boundary into the interior. Such entropy-regularised dynamics connect to the perturbed best-response (logit) dynamics of smooth fictitious play \cite{Fudenberg1998TheoryLearning, Hofbauer2002GlobalConvergence}, whose interior rest points are the quantal response equilibria of \citeauthor{McKelvey1995QuantalResponse}~\cite{McKelvey1995QuantalResponse}. The added negative entropy term also breaks the exact orthogonal decoupling of \S\ref{sec:topological-equivalence}, since it introduces a non-zero self-interaction curvature $\partial^2 F / \partial x_1^2 \ne 0$.

Subsequent work must also extend the pairwise framework to higher-order, $N$-way strategic interactions and to games beyond two actions ($n > 2$). The inversion of pure-strategy dominance itself requires no such extension, since Proposition~\ref{prop:dominance-inversion} already covers every $n \times m$ bimatrix game. The complete topological classification, by contrast, does need extension. With more than two state coordinates, the planar arguments of \S\ref{sec:analysis-2x2-games} no longer apply. The zero trace survives for $N > 2$ two-action agents (\S\ref{sec:hamiltonian-structure}), but global topological guarantees will need new analytical methods. Tensor payoffs also enter the two models differently: under linear fitness they make each agent's fitness a multilinear polynomial in the opponents' mixed strategies, whereas under log-linear fitness their effect depends on how \eqref{eqn:log-linear-fitness} is extended to $N$ players.

\appendix

\section{Proof of Theorem~\ref{thm:partition}}\label{apx:partition-proof}

This appendix proves the partition theorem of \S\ref{sec:topological-equivalence}. Notation is that of the theorem; in addition, for an interior state $z$ and a phase flow $\phi$, let $T^{\pm}(z, \phi) \subseteq \bar{\mathcal{M}}$ denote the non-empty closed set of limit points of $\phi(z, t)$ as $t \to \pm\infty$. The necessity argument rests on two elementary properties of vertex-preserving equivalences.

\begin{lemma}[Limit transport under vertex-preserving equivalence]\label{lem:limit-transport}
Let $h$ be a topological equivalence from $(\mathcal{M}, \mathcal{X}_{\mathrm{Log}})$ to $(\mathcal{M}, \mathcal{X}_{\mathrm{Lin}})$ admitting a continuous extension $\bar{h}$ as in Definition~\ref{def:topological-equivalence}. Then:
\begin{enumerate}[label=(\roman*)]
    \item $\bar{h}$ maps boundary points of $\bar{\mathcal{M}}$ to boundary points;
    \item if $p \in T^{+}(z, \phi_\mathrm{Log})$ is a boundary point, then $\bar{h}(p) \in T^{+}(h(z), \phi_\mathrm{Lin})$, and likewise for $T^{-}$.
\end{enumerate}
\end{lemma}

\begin{proof}
(i) Suppose $\bar{h}(p) = h(q)$ for a boundary point $p$ and an interior point $q$. Choosing interior states $z_n \to p$ gives $h(z_n) \to h(q)$, and continuity of $h^{-1}$ forces $z_n \to q \ne p$, a contradiction.

(ii) Take $t_k \to \infty$ with $\phi_\mathrm{Log}(z, t_k) \to p$. Then $h(\phi_\mathrm{Log}(z, t_k)) = \phi_\mathrm{Lin}(h(z), \tau(z, t_k)) \to \bar{h}(p)$. The increasing sequence $\tau(z, t_k)$ cannot converge to a finite limit $\bar{\tau}$: this would make $\bar{h}(p) = \phi_\mathrm{Lin}(h(z), \bar{\tau})$ an interior point, contradicting (i). Hence $\tau(z, t_k) \to \infty$, and $\bar{h}(p)$ is a limit point of the linear orbit.
\end{proof}

\begin{proof}[Proof of Theorem~\ref{thm:partition}]
\textbf{Part I: Necessity.} We first show that if some agent satisfies $\mathcal{P}_i$, no vertex-preserving topological equivalence exists. As a first indication, compare the orientations of the two vector fields near the absolute boundaries ($y_2 \to 0$ and $y_2 \to 1$).

In the linear model, evaluating the scalar velocity $\dot{v}_\mathrm{Lin} = -\Delta_0^i y_1 - \Delta_1^i y_2$ at the boundaries ($y_2 \to 0 \implies y_1 \to 1$) yields direct algebraic limits:
\begin{align*}
\lim_{y_2 \to 0} \operatorname{sign}(\dot{v}_\mathrm{Lin}) &= \operatorname{sign}(-\Delta_0^i) \\
\lim_{y_2 \to 1} \operatorname{sign}(\dot{v}_\mathrm{Lin}) &= \operatorname{sign}(-\Delta_1^i)
\end{align*}

In the log-linear model, the scalar velocity $\dot{v}_\mathrm{Log} = -\Delta_0^i \ln y_1 - \Delta_1^i \ln y_2$ diverges logarithmically at the boundaries. However, because topological orientation depends strictly on the sign of the velocity, we can extract the limiting orientations directly without requiring time-rescaling or compactification. 
As $y_2 \to 0$, $y_1 \to 1$, which forces $\ln y_1 \to 0$ and $\ln y_2 \to -\infty$: the velocity is dominated by the term $-\Delta_1^i \ln y_2$, which diverges with the sign of $\Delta_1^i$ (and vanishes with it when $\Delta_1^i = 0$). Symmetrically, as $y_2 \to 1$ the term $-\Delta_0^i \ln y_1$ dominates. Hence
\begin{align*}
\lim_{y_2 \to 0} \operatorname{sign}(\dot{v}_\mathrm{Log}) &= \operatorname{sign}(\Delta_1^i) \\
\lim_{y_2 \to 1} \operatorname{sign}(\dot{v}_\mathrm{Log}) &= \operatorname{sign}(\Delta_0^i)
\end{align*}

Because the log-linear vector field swaps the mapping of the $\Delta_0^i$ and $\Delta_1^i$ terms to the respective boundaries, the boundary orientations of the two flows can only agree when $\Delta_0^i$ and $\Delta_1^i$ have opposite signs; their agreement in sign, or the vanishing of one but not the other, is Condition $\mathcal{P}_i$ \eqref{eqn:preservation-condition}. Boundary orientations alone, however, do not rule out a homeomorphism. The argument proceeds in four stages: the asymptotic opposition is upgraded to a global one; the possible limit sets of both flows are confined to closed boundary sets, the \emph{corridors} of the argument; Lemma~\ref{lem:limit-transport} then yields two mechanisms that force a vertex off itself; and a case analysis confirms that one of the two mechanisms is always available.

\emph{Global sign opposition.} On every branch of $\mathcal{P}_i$, the two scalar velocities have constant, opposite signs on the entire interior:
\begin{itemize}
    \item if $\Delta_0^i, \Delta_1^i > 0$, then $\dot{v}_\mathrm{Log}$ is a sum of two positive terms (every $\ln y_k < 0$ on the open simplex) whilst $\dot{v}_\mathrm{Lin}$ is a sum of two negative terms, so $\dot{v}_\mathrm{Log} > 0 > \dot{v}_\mathrm{Lin}$ everywhere;
    \item if $\Delta_0^i, \Delta_1^i < 0$, then symmetrically $\dot{v}_\mathrm{Log} < 0 < \dot{v}_\mathrm{Lin}$ everywhere;
    \item if $\Delta_0^i = 0 \ne \Delta_1^i$, then $\dot{v}_\mathrm{Log} = -\Delta_1^i \ln y_2$ carries the sign of $\Delta_1^i$ whilst $\dot{v}_\mathrm{Lin} = -\Delta_1^i y_2$ carries the sign of $-\Delta_1^i$ (symmetrically when $\Delta_1^i = 0 \ne \Delta_0^i$).
\end{itemize}
Games with equal non-zero incentives ($\Delta_0^i = \Delta_1^i \ne 0$) fall under the first two branches. Relabelling strategies or agents as necessary, we may assume the agent satisfying $\mathcal{P}_i$ is Agent A, with $\dot{v}_\mathrm{Log} > 0 > \dot{v}_\mathrm{Lin}$ on all of $\mathcal{M}$; relabelling Agent B's strategies if necessary (which exchanges $\Delta_0^A$ with $\Delta_1^A$ and negates Agent B's own differentials, over which the case analysis below is exhaustive), we may further assume $\Delta_0^A = 0 < \Delta_1^A$ in any half-degenerate case. The corridor derivation next uses this orientation. Suppose, for contradiction, that $h$ is a topological equivalence with continuous vertex-fixing extension $\bar{h}$, and describe boundary points of $\bar{\mathcal{M}}$ by their $(x_1, y_1)$ coordinates, writing $F_1 = \{x_1 = 1\}$, $F_0 = \{x_1 = 0\}$, $G_1 = \{y_1 = 1\}$ and $G_0 = \{y_1 = 0\}$ for the boundary faces.

\emph{Limit corridors.} Along every log-linear orbit $v$ is increasing, so either $v \to +\infty$ and the forward limit points lie on $F_1$, or $v$ converges (the orbit \emph{stalls}), and then $y_1(t) \to 0$: the forward limit points lie on $G_0$. The stalling case rests on two bounds. First, by the branch formulae, $\dot{v}_\mathrm{Log}$ is bounded below by a positive constant wherever $y_1 \ge \varepsilon$, so the finite total increment of $v$ bounds the total time spent there. Second, the opponent's velocity is bounded along the tail, since the stalled coordinate confines $x$ to a compact interior interval, so each excursion of $y_1$ above $\varepsilon$ spends a uniformly positive time above $\varepsilon/2$, whilst the first bound makes the total time above $\varepsilon/2$ finite: excursions eventually cease, and $y_1(t) \to 0$. Thus
$$T^{+}(z, \phi_\mathrm{Log}) \subseteq F_1 \cup G_0,$$
and in particular every limit point is a boundary point, so Lemma~\ref{lem:limit-transport}(ii) applies to it. Symmetrically, linear orbits stall only towards $y_1 \to 1$, so
$$T^{+}(h(z), \phi_\mathrm{Lin}) \subseteq F_0 \cup G_1,$$
a closed set that does not contain the vertex $(1, 0)$. Because $v$ decreases along linear orbits, we also have $T^{+}(h(z), \phi_\mathrm{Lin}) \subseteq \{x_1 \le x_1(h(z))\}$ and $T^{-}(h(z), \phi_\mathrm{Lin}) \subseteq \{x_1 \ge x_1(h(z))\}$: no forward linear limit point lies on $F_1$, and no backward one on $F_0$.

\emph{Contradiction mechanisms.} Combining the corridors with Lemma~\ref{lem:limit-transport} gives two routes to a contradiction.
\begin{itemize}
    \item \emph{Vertex limit.} If some log-linear orbit converges forwards to a vertex $V \in F_1$ (or backwards to a vertex $V \in F_0$), then $\bar{h}(V) = V$ lies in a set containing no point of $F_1$ (respectively $F_0$), which is impossible.
    \item \emph{Sweep.} If forward limit points of a family of log-linear orbits accumulate at a vertex $V$, their $\bar{h}$-images lie in the closed corridor $F_0 \cup G_1$ (in several cases below, in the smaller closed face $F_0$), and continuity of $\bar{h}$ then places $\bar{h}(V) = V$ in that closed set, which is impossible whenever the set excludes $V$.
\end{itemize}

\emph{Case analysis.} It remains to check that one of the two mechanisms is always available. Write $\dot{w} = -\Delta_0^B \ln x_1 - \Delta_1^B \ln x_2$ for the opponent's log-linear velocity.

On the main branch ($\Delta_0^A \Delta_1^A > 0$), the positive function $-\Delta_0^A \ln y_1 - \Delta_1^A \ln y_2$ diverges at both endpoints of the opponent's simplex and is therefore bounded below by a positive constant: no log-linear orbit stalls, $v$ grows at least linearly, and $x_2$ decays exponentially along every orbit. Likewise $|\dot{v}_\mathrm{Lin}| \ge \min(\Delta_0^A, \Delta_1^A) > 0$, so linear orbits do not stall either and $T^{+}(\cdot, \phi_\mathrm{Lin}) \subseteq F_0$.
\begin{itemize}
    \item \emph{$\Delta_1^B \ne 0$.} As $x_1 \to 1$ the opponent's velocity $\dot{w} \to -\Delta_1^B \ln x_2$ diverges with the fixed sign of $\Delta_1^B$, so $w \to \pm\infty$ and every log-linear orbit converges to a vertex of $F_1$: vertex limit.
    \item \emph{$\Delta_1^B = 0$.} Now $|\dot{w}| = |\Delta_0^B|\,|\ln x_1|$ decays exponentially along each orbit, $w$ converges, and the orbit converges to a point $(1, y_\infty) \in F_1$ whose logit differs from the initial one by a uniformly bounded amount. Letting the initial logit diverge, the limit points accumulate at a vertex of $F_1$: sweep, with the closed face $F_0$.
\end{itemize}

On the half-degenerate branch ($\Delta_0^A = 0 < \Delta_1^A$ under the chosen orientation; the mirrored variants follow by relabelling), $\dot{v}_\mathrm{Log} = \Delta_1^A |\ln y_2|$ degenerates as $y_1 \to 0$, and a stall at the boundary point $(\bar{x}_1, 0)$ requires $w \to -\infty$ there, hence a non-positive limiting opponent drift $c(\bar{v}) = -\Delta_0^B \ln \bar{x}_1 - \Delta_1^B \ln \bar{x}_2 \le 0$. We distinguish the opponent's parameters.
\begin{itemize}
    \item \emph{$\Delta_1^B > 0$.} Here $c(v) \to +\infty$ as $v \to +\infty$, so orbits started at sufficiently large $v$ cannot stall; along them $v \to \infty$, then $\dot{w} \to -\Delta_1^B \ln x_2 \to +\infty$, so $w \to +\infty$ and the orbit converges to the vertex $(1, 1)$: vertex limit.
    \item \emph{$\Delta_1^B < 0$.} Every orbit stalls: were $v \to \infty$, eventually $\dot{w} \le -1$, so $w \to -\infty$ at least linearly, $y_1 \sim e^{w}$ would decay exponentially, and $\dot{v}_\mathrm{Log} = \Delta_1^A |\ln(1 - y_1)| \sim \Delta_1^A y_1$ would be integrable, forcing $v$ to converge. The stall values $\bar{v}(z) > v(z)$ are therefore unbounded above across orbits, and the stall points $(\sigma(\bar{v}), 0) \in G_0$ accumulate at the vertex $(1, 0)$, which the closed corridor $F_0 \cup G_1$ excludes: sweep.
    \item \emph{$\Delta_1^B = 0 = \Delta_0^B$.} Agent B's velocity vanishes identically and $\dot{v}_\mathrm{Log}$ is constant in time along each orbit, so no orbit stalls and the forward limit points $(1, y_0)$ sweep the entire face $F_1$; the linear flow never stalls either ($\dot{v}_\mathrm{Lin}$ is a non-zero constant along each orbit): sweep, with $F_0$.
    \item \emph{$\Delta_1^B = 0 \ne \Delta_0^B$.} The opponent's velocity $\dot{w} = -\Delta_0^B \ln x_1$ has a fixed sign. For $\Delta_0^B > 0$, $w$ increases, $y_2$ decreases, $\dot{v}_\mathrm{Log}$ is bounded below along each orbit, no stalling occurs, $w$ converges (exponential decay of $|\ln x_1|$ as in the main branch), and the forward limit points $(1, y_\infty)$ accumulate at the vertex $(1, 1)$; the linear flow cannot stall here (a linear stall requires $y_1 \to 1$, but the opponent's linear velocity $-\Delta_0^B x_1$ drives $y_1$ downwards), so $T^{+}(\cdot, \phi_\mathrm{Lin}) \subseteq F_0$: sweep. For $\Delta_0^B < 0$, argue backwards in time: backwards, $v$ decreases and $w$ increases, so $y_2$ decreases, $\dot{v}_\mathrm{Log}$ stays bounded below, no backward stall occurs, $v \to -\infty$, and $|\dot{w}| = |\Delta_0^B|\,|\ln x_1| \to \infty$ sends $w \to +\infty$: every orbit converges backwards to the vertex $(0, 1) \in F_0$, and the vertex-limit mechanism applies in backward time.
\end{itemize}
In every case the assumed equivalence is contradicted. Because the two-dimensional flow decouples orthogonally, it suffices for Condition $\mathcal{P}_i$ to hold for a single agent, so topological non-equivalence covers the whole of $\mathcal{D}$.

\medskip
\textbf{Part II: Sufficiency.} We now show that on the complement $\mathcal{E}$, boundary sign agreement dictates identical global phase portraits. An agent in $\mathcal{E}$ either satisfies $\Delta_0^i \Delta_1^i < 0$ or is fully degenerate ($\Delta_0^i = \Delta_1^i = 0$).

\textbf{Step 0: Degenerate agents.} If an agent is fully degenerate, its velocity vanishes identically under both models, and the joint dynamics reduce to a family of one-dimensional flows in the opponent's coordinate with the degenerate agent's coordinate constant. If the opponent is also degenerate, both joint fields are identically zero and the flows are trivially equivalent. Otherwise the opponent satisfies $\Delta_0 \Delta_1 < 0$, and by the argument of Step 1 below its velocity has the same unique sign change under both models; a skew-product homeomorphism $h(v, w) = (\alpha(v), \beta(v, w))$ preserving the invariant direction realises the equivalence, where $\alpha$ is an increasing homeomorphism carrying the opponent's log-linear rest line to its linear rest line (the two rest lines sit at different values of the frozen coordinate, so that coordinate cannot be fixed pointwise), and $\beta(v, \cdot)$ matches the resulting one-dimensional flows, whose orientations agree on each side of the rest line. For the remainder of the proof we may therefore assume both agents satisfy $\Delta_0^i \cdot \Delta_1^i < 0$ (the boundary incentives have strictly opposite signs).

\textbf{Step 1: Unique Interior Equilibrium.} In the log-linear model, an interior nullcline requires $\dot{v}_\mathrm{Log} = -\Delta_0^i \ln y_1 - \Delta_1^i \ln y_2 = 0$. Because $\ln y_1, \ln y_2 < 0$ on the open simplex, their weighted sum has a strict, non-zero sign everywhere if $\Delta_0^i \Delta_1^i > 0$, meaning no interior zero exists. Conversely, if $\Delta_0^i \Delta_1^i < 0$, the partial derivative $\frac{\partial \dot{v}_\mathrm{Log}}{\partial y_1} = -\frac{\Delta_0^i}{y_1} + \frac{\Delta_1^i}{y_2}$ maintains a constant sign on $(0, 1)$, rendering $\dot{v}_\mathrm{Log}$ strictly monotone with limits $\mp \infty$, which yields exactly one interior zero. Similarly, the linear model has an interior zero at $y_1^\mathrm{Lin} = \frac{\Delta_1^i}{\Delta_1^i - \Delta_0^i}$, which lies in the open interval $(0, 1)$ if and only if $\Delta_0^i \Delta_1^i < 0$. Thus, both models contain exactly one interior nullcline for agent $i$ if and only if $\Delta_0^i \Delta_1^i < 0$.

\textbf{Step 2: Zero Trace Structure.} In both models, the 2D planar system is orthogonally decoupled: Agent A's velocity $\dot{v}$ depends only on the opponent's state $w$, and Agent B's velocity $\dot{w}$ depends only on $v$. Therefore, the diagonal entries of the Jacobian matrix are identically zero ($\partial \dot{v}/\partial v = \partial \dot{w}/\partial w = 0$), so the trace vanishes ($\nabla \cdot \mathcal{X} = 0$).

\textbf{Step 3: Matching Off-Diagonal Jacobian Signs.} Because the trace is zero, the local stability classification is determined by the determinant $\det(J) = -(\partial \dot{v}/\partial w)(\partial \dot{w}/\partial v)$. By applying the chain rule $\frac{\partial \dot{v}}{\partial w} = \frac{\partial \dot{v}}{\partial y_1} \frac{\partial y_1}{\partial w}$ and noting that the mapping is strictly monotonic ($\frac{\partial y_1}{\partial w} = y_1 y_2 > 0$), we can determine the sign of the log-odds derivative directly from the probability-coordinate derivative $\partial \dot{v}/\partial y_1$. In the linear model, this derivative evaluates to $\partial \dot{v}/\partial y_1 = -\Delta_0 + \Delta_1$. In the log-linear model, it evaluates to $\partial \dot{v}/\partial y_1 = -\frac{\Delta_0}{y_1} + \frac{\Delta_1}{y_2}$. Because $\Delta_0$ and $\Delta_1$ have opposite signs, the two derivatives share the same sign. Therefore, the sign of $\det(J)$ matches between the two models: if the linear model has a saddle ($\det < 0$), the log-linear model has a saddle; if the linear model has a centre ($\det > 0$), the log-linear model has a centre.

\textbf{Step 4: Global Orbit Structure via Energy Decomposition.} Both flows admit an elementary first integral built from the scalar velocities of \S\ref{sec:topological-equivalence} alone. Because each velocity depends only on the opponent's coordinate, we may define $g_A$ as an antiderivative of $w \mapsto \dot{v}_\mathrm{Log}(w)$ and $g_B$ as an antiderivative of $v \mapsto \dot{w}_\mathrm{Log}(v)$, and set
$$H(v, w) = g_A(w) - g_B(v).$$
Along any trajectory, $\dot{H} = g_A'(w)\,\dot{w} - g_B'(v)\,\dot{v} = \dot{v}\dot{w} - \dot{w}\dot{v} = 0$, so $H$ is conserved and every orbit lies on one of its level sets. (\S\ref{sec:hamiltonian-structure} develops this first integral into the full Hamiltonian structure and gives $g_A$, $g_B$ in closed form.) Differentiating the velocity with the chain rule of Step 3 ($\partial y_1 / \partial w = y_1 y_2$) gives
$$g_A''(w) = \frac{d \dot{v}_\mathrm{Log}}{d w} = \left( -\frac{\Delta_0^A}{y_1} + \frac{\Delta_1^A}{y_2} \right) y_1 y_2 = \Delta_1^A y_1 - \Delta_0^A y_2.$$
Because $\Delta_0^A \Delta_1^A < 0$, this expression has a constant sign: $g_A$ is strictly convex for a coordination-type agent ($\Delta_0^A < 0 < \Delta_1^A$) and strictly concave for an anti-coordination type. Moreover $g_A$ is coercive: since $\ln y_2 \sim -w$ as $w \to +\infty$ and $\ln y_1 \sim w$ as $w \to -\infty$, the velocity grows linearly, $\dot{v}_\mathrm{Log} \sim \Delta_1^A w$ and $\sim -\Delta_0^A w$ respectively, so $g_A(w) \sim \tfrac{1}{2}\Delta_1^A w^2$ as $w \to +\infty$ and $g_A(w) \sim -\tfrac{1}{2}\Delta_0^A w^2$ as $w \to -\infty$; for a coordination type both coefficients are positive (for anti-coordination, both negative). The same holds for $g_B$. The linear flow is treated identically, with $h_A$, $h_B$ defined as antiderivatives of the linear velocities and $H_\mathrm{Lin} = h_A(w) - h_B(v)$ conserved by the same computation: $h_A''(w) = (\Delta_1^A - \Delta_0^A)\, y_1 y_2$ carries the same constant sign as $g_A''$ on $\mathcal{E}$, whilst $h_A'(w) = \dot{v}_\mathrm{Lin} \to -\Delta_0^A$ and $-\Delta_1^A$ as $w \to +\infty$ and $-\infty$, so $h_A$ is likewise strictly convex (or concave) and coercive. Two global orbit structures follow, identically in both models.

\emph{Mixed coordination types (centre).} If $g_A$ is strictly convex and $g_B$ strictly concave (or vice versa), then $H = g_A(w) + (-g_B)(v)$ is, up to overall sign, a sum of strictly convex coercive functions of separate variables: it is strictly convex and coercive with the interior equilibrium of Step 1 as its unique critical point, and every level set above the minimum is a simple closed convex curve encircling it. Orbits coincide with these level curves, so the flow is a global centre with no limit cycles.

\emph{Matching coordination types (saddle).} If $g_A$ and $g_B$ are both strictly convex (both concave is analogous), the critical level set through the equilibrium, $\{g_A(w) - g_A(w^*) = g_B(v) - g_B(v^*)\}$, consists of exactly four separatrix branches: for each $v \ne v^*$ the strictly convex, coercive function $w \mapsto g_A(w) - g_A(w^*)$ attains the positive value $g_B(v) - g_B(v^*)$ at exactly two points, so the set is a pair of curves crossing only at the equilibrium. The quadrant signs of $(\dot{v}, \dot{w}) = (g_A'(w), g_B'(v))$ send the two unstable branches to the compactified corners $(v, w) \to (+\infty, +\infty)$ and $(-\infty, -\infty)$, whilst the two stable branches emanate from $(+\infty, -\infty)$ and $(-\infty, +\infty)$. The same computation for $h_A, h_B$ assigns the linear separatrices to the same four corners.

\textbf{Step 5: Explicit Vertex-Fixing Equivalence.} Equilibrium types, corner assignments, and boundary orientations now agree, and we construct the homeomorphism directly, quadrant by quadrant, in energy coordinates. Let $(v^*_\mathrm{Log}, w^*_\mathrm{Log})$ and $(v^*_\mathrm{Lin}, w^*_\mathrm{Lin})$ denote the two interior equilibria (Step 1), and define
\begin{gather*}
    p_\mathrm{Log}(v) = \big|g_B(v) - g_B(v^*_\mathrm{Log})\big|, \\
    q_\mathrm{Log}(w) = \big|g_A(w) - g_A(w^*_\mathrm{Log})\big|,
\end{gather*}
with $p_\mathrm{Lin}, q_\mathrm{Lin}$ built analogously from $h_B, h_A$. Each of these four functions is continuous, vanishes only at the corresponding equilibrium coordinate, is strictly monotone on each side of it, and diverges at $\pm\infty$ (strict convexity or concavity plus coercivity, Step 4); hence $(v, w) \mapsto (p, q)$ restricts to a homeomorphism from each closed quadrant of the nullcline grid onto $[0, \infty)^2$, in both models. Define $h$ on each closed quadrant as the energy-coordinate transport
$$h \triangleq (p_\mathrm{Lin}, q_\mathrm{Lin})^{-1} \circ (p_\mathrm{Log}, q_\mathrm{Log}),$$
matching quadrants by their sign pattern relative to the respective nullclines.

The four quadrant maps agree along the shared nullcline half-lines (where $p = 0$ or $q = 0$) and at the equilibrium, so $h$ is a homeomorphism of the plane sending equilibrium to equilibrium. It maps orbits onto orbits: in energy coordinates the conserved quantity reads $H - H^{*} = q + p$ in the centre case and $H - H^{*} = q - p$ in the saddle case, with the same form in both models, so each log-linear level curve is carried onto the linear level curve of the same energy offset, component by component; in particular, separatrices map to separatrices. Time orientation is preserved because within each quadrant the signs of $(\dot{v}, \dot{w})$ are determined by the quadrant alone and agree between the two models (Steps 3--4).

The reparametrisation $\tau$ demanded by Definition~\ref{def:topological-equivalence} is explicit in the same coordinates. Since $h$ is the identity in $(p, q)$, both flows traverse the same $(p, q)$-path. Along the log-linear flow $|\dot{p}| = |\dot{q}| = |g_A'(w)\, g_B'(v)|$, since $\dot{v} = g_A'(w)$ and $\dot{w} = g_B'(v)$, and likewise for the linear flow with $h_A, h_B$; the ratio of the two speeds at corresponding points $(v, w)$ and $(\tilde{v}, \tilde{w}) = h(v, w)$ factorises as $R(v, w) = R_B(v)\, R_A(w)$ with $R_A(w) = |g_A'(w)| \,/\, |h_A'(\tilde{w}(w))|$. Each factor extends to a continuous strictly positive function on all of $\mathbb{R}$: away from the nullcline value it is a composition of continuous non-vanishing maps, and as $w \to w^*_\mathrm{Log}$ numerator and denominator both vanish like the square root of the common energy $q$, with ratio tending to $\sqrt{|g_A''(w^*_\mathrm{Log})| / |h_A''(w^*_\mathrm{Lin})|}$, the second derivatives being non-zero on $\mathcal{E}$ (Step 4). Setting $\tau(z, t) = \int_0^t R\big(\phi_\mathrm{Log}(z, s)\big)\, ds$ then gives a jointly continuous function, strictly increasing in $t$ with $\tau(z, 0) = 0$, and $h(\phi_\mathrm{Log}(z, t)) = \phi_\mathrm{Lin}(h(z), \tau(z, t))$ holds because both sides start at $h(z)$ and traverse the same $(p, q)$-path at equal speeds under this time change.

Finally, coercivity extends each coordinate map to a homeomorphism of the compactified quadrant $[0, \infty]^2$, so $h$ extends continuously to $\bar{h} : \bar{\mathcal{M}} \to \bar{\mathcal{M}}$, carrying each boundary face onto itself and fixing every vertex, the vertices being the outer corner points of the four compactified quadrants. This gives the vertex-preserving topological equivalence required by Definition~\ref{def:topological-equivalence}.\footnote{We construct $h$ explicitly because the classical route through the Markus--Neumann classification of planar flows by their separatrix configurations is unavailable here for two reasons: the classification is false under the notion of separatrix originally employed, as shown by the counterexamples of \citeauthor{EspinBuendia2018OnMarkusNeumann}~\cite{EspinBuendia2018OnMarkusNeumann} (see \citeauthor{Braun2025RevisitingMarkusNeumann}~\cite{Braun2025RevisitingMarkusNeumann} for a corrected statement with a self-contained proof); and even in its corrected form it yields an unconstrained homeomorphism with no control at the boundary, whereas Definition~\ref{def:topological-equivalence} demands a vertex-fixing boundary extension.}
Together with Step 0, this completes the proof that $\mathcal{E}$ contains only topologically equivalent games.

\medskip
\textbf{Part III: Characterisations of $\mathcal{D}$.} The two descriptions of $\mathcal{D}$ in the statement follow directly. Strategy 2 weakly dominates Strategy 1 for agent $i$ if and only if $T_i \ge R_i$ and $P_i \ge S_i$, i.e. $\Delta_0^i \ge 0$ and $\Delta_1^i \ge 0$, with strict dominance when both inequalities are strict; symmetrically, Strategy 1 weakly dominates when both differentials are non-positive. Hence $\mathcal{P}_i$ holds if and only if agent $i$ possesses a weakly dominant pure strategy and is not fully indifferent. For the rest-point characterisation: agent $i$'s velocity either vanishes identically (full indifference) or possesses an interior zero if and only if $\Delta_0^i \Delta_1^i < 0$ (Step 1, in both models); a joint interior rest point therefore exists if and only if every agent fails $\mathcal{P}_i$, i.e. if and only if the game lies in $\mathcal{E}$.
\end{proof}

\section*{CRediT authorship contribution statement}
\textbf{\authorRuizSerra:} Conceptualization (lead); Methodology (lead); Formal analysis (lead); Investigation (lead); Software (lead); Visualization (lead); Writing -- original draft (lead); Writing -- review \& editing (lead).

\textbf{\authorAy:} Conceptualization (supporting); Writing -- review \& editing (supporting).

\textbf{\authorHarre:} Conceptualization (supporting); Supervision (lead); Writing -- review \& editing (supporting).

\section*{Declaration of competing interest}
The authors declare that they have no known competing financial interests or personal relationships that could have appeared to influence the work reported in this paper.

\section*{Data availability}
No data was used for the research described in the article. The script that draws the phase portraits (Figures~\ref{fig:sh-displacement}--\ref{fig:mp-dual}) is available from the corresponding author on request.

\section*{Funding}
This work was supported by an Australian Government Research Training Program (RTP) Scholarship. The funder had no role in the study design, the analysis, the writing of the article, or the decision to submit it for publication.

\section*{Acknowledgements}
We acknowledge the Gadi people of the Eora Nation as the traditional custodians of the land on which The University of Sydney now stands, and that their sovereignty was never ceded.

\section*{Declaration of generative AI and AI-assisted technologies in the manuscript preparation process}
During the preparation of this work the authors used the generative AI models Gemini Pro 3.8, Claude Fable 5.1 and Claude Opus 5.5 for drafting and adversarially checking candidate proofs, for writing the script that draws the phase portraits, and for copy editing. After using these tools, the authors reviewed and edited the content as needed and take full responsibility for the content of this article, including all AI-assisted material.
 
\bibliography{references-bibtex}

\end{document}